\documentclass[structabstract,printer]{aa}  
\usepackage{graphicx}
\usepackage{lscape}
\usepackage{longtable}
\usepackage{natbib}
\usepackage{tikz}
\usetikzlibrary{shapes,arrows}
\usepackage[varg]{txfonts}
\usepackage{csquotes}
\MakeOuterQuote{"}

\usetikzlibrary{shapes.misc}

\newcommand{\Teff}{{\mathrm{T_{eff}}}}
\newcommand{\FeH}{{\mathrm{[Fe/H]}}}
\newcommand{\logg}{\log g}

\begin{document}

\title{The Gaia-ESO Survey: Extracting diffuse interstellar bands from cool star spectra
\thanks{Based on observations made with the
ESO/VLT at Paranal Observatory, under programs
188.B-3002 (The Gaia-ESO Public Spectroscopic Survey) and 079.B-0662.}
}
\subtitle{DIB-based interstellar medium line-of-sight structures at the kpc scale}

\author{L. Puspitarini\inst{1}, R. Lallement\inst{1}, C. Babusiaux\inst{1}, H-C. Chen\inst{2}, P. Bonifacio\inst{1}, L. Sbordone\inst{3}, E. Caffau\inst{1}, S. Duffau\inst{3}, V. Hill\inst{4}, A. Monreal-Ibero\inst{1}, F. Royer\inst{1}, F. Arenou\inst{1}, R., A. Peralta\inst{5}, J.E. Drew\inst{6}, R. Bonito\inst{7}, J. Lopez-Santiago\inst{8}, E. Alfaro\inst{9}, T. Bensby\inst{10}, A. Bragaglia\inst{11}, E. Flaccomio\inst{7}, A. Lanzafame\inst{12}, E. Pancino\inst{11}, A. Recio-Blanco\inst{4}, R. Smiljanic\inst{13}, M.T. Costado\inst{9}, C. Lardo\inst{11}, P. de Laverny\inst{4}, T. Zwitter\inst{14}}
\institute{GEPI, Observatoire de Paris, CNRS UMR8111, Universit\'e Paris Diderot, Place Jules Janssen,  92190 Meudon, France\\
\email{lucky.puspitarini@obspm.fr, rosine.lallement@obspm.fr},
\and{Institute of Astronomy, National Central  University, Chungli, Taiwan}
\and{Zentrum f\"ur Astronomie der Universit\"at Heidelberg, Landessternwarte, K\"onigstuhl 12, 69117, Heidelberg, Germany, Millennium Institute of Astrophysics, Pontificia Universidad Cat\'olica de Chile, Av. Vicuna Mackenna 4860, 782-0436 Macul, Santiago, Chile}
\and{Laboratoire Lagrange (UMR7293), Universit\'e de Nice Sophia Antipolis, CNRS, Observatoire de la Cote d'Azur, bd. de l'Observatoire, BP 4229, 06304, Nice Cedex 4, France}
\and{LESIA, Observatoire de Paris, CNRS UMR8109,  l'Universit\'e Pierre et Marie Curie (Paris 6), 5 Place Jules Janssen,  92190 Meudon, France}
\and{Centre for Astrophysics Research, STRI, University of Hertfordshire, College Lane Campus, Hatfield AL10 9AB, United Kingdom}
\and{INAF - Osservatorio Astronomico di Palermo, Piazza del Parlamento 1, and Dipartimento di Fisica e Chimica, Universita' di Palermo, 90134, Palermo, Italy}
\and{Univ. Madrid, Departamento de Astrofisica, Spain}
\and{Instituto de Astrof\'{i}sica de Andaluc\'{i}a-CSIC, Apdo. 3004, 18080, Granada, Spain}
\and{Lund Observatory, Department of Astronomy and Theoretical Physics, Box 43, SE-221 00 Lund, Sweden}
\and{INAF - Osservatorio Astronomico di Bologna, via Ranzani 1, 40127, Bologna, Italy}
\and{Dipartimento di Fisica e Astronomia, Sezione Astrofisica, Universit\`{a} di Catania, via S. Sofia 78, 95123, Catania, Italy}
\and{Department for Astrophysics, Nicolaus Copernicus Astronomical Center, ul. Rabia\'nska 8, 87-100 Toru\'n, Poland}
\and{Faculty of Mathematics and Physics, University of Ljubljana, Jadranska 19, 1000 Ljubljana, Slovenia.}}


\abstract
{}
{We study how diffuse interstellar bands (DIBs) measured toward distance-distributed target stars can be used to locate dense interstellar (IS) clouds in the Galaxy and probe a line-of-sight (LOS) kinematical structure, a potential useful tool when gaseous absorption lines are saturated or not available in the spectral range. Cool target stars are numerous enough for this purpose.}
{We have devised automated DIB fitting methods appropriate to cool star spectra and multiple IS components. The data is fitted with a combination of a synthetic stellar spectrum, a synthetic telluric transmission, and empirical DIB profiles. The initial number of DIB components and their radial velocity 
are guided by HI 21 cm emission spectra, or, when available in the spectral range, IS neutral sodium absorption lines. In case of NaI, radial velocities of NaI lines and DIBs are kept linked during a global simultaneous fit. In parallel, stellar distances and extinctions are estimated self-consistently by means of a 2D Bayesian method, from spectroscopically-derived stellar parameters and photometric data.}
{We have analyzed Gaia-ESO Survey (GES) spectra of 225 stars that probe between $\sim$ 2 and 10 kpc long LOS in five different regions of the Milky Way. The targets are the two CoRoT fields, two open clusters (NGC4815 and $\gamma$ Vel), and the Galactic bulge. Two OGLE fields toward the bulge observed prior to GES are also included (205 target stars). Depending on the observed spectral intervals, we extracted one or more of the following DIBs: $\lambda\lambda$ 6283.8, 6613.6 and 8620.4.  For each field, we compared the DIB strengths with the Bayesian distances and extinctions, and the DIB Doppler velocities with the HI emission spectra.}
{For all fields, the DIB strength and the target extinction are well correlated. In case of targets widely distributed in distance, marked steps in DIBs and extinction radial distance profiles match with each other and broadly correspond to the expected locations of spiral arms. For all fields, the DIB velocity structure agrees with HI emission spectra and all detected DIBs correspond to strong NaI lines. This illustrates how DIBs can be used to locate the Galactic interstellar gas and to study its kinematics at the kpc scale, as illustrated by Local and Perseus Arm DIBs that differ by $\gtrsim$ 30 km/s, in agreement with HI emission spectra. On the other hand,  if most targets are located beyond the main absorber, DIBs can trace the differential reddening within the field.}
\keywords{ISM; diffuse interstellar band (DIB); extinction; color excess; DIB-distance; Perseus arm}

\authorrunning{Puspitarini et al}
\titlerunning{Extraction of Multi-Component Diffuse Interstellar Bands}
\maketitle

\section{Introduction}

\begin{figure*}[t]
\centering
\includegraphics[width=\textwidth]{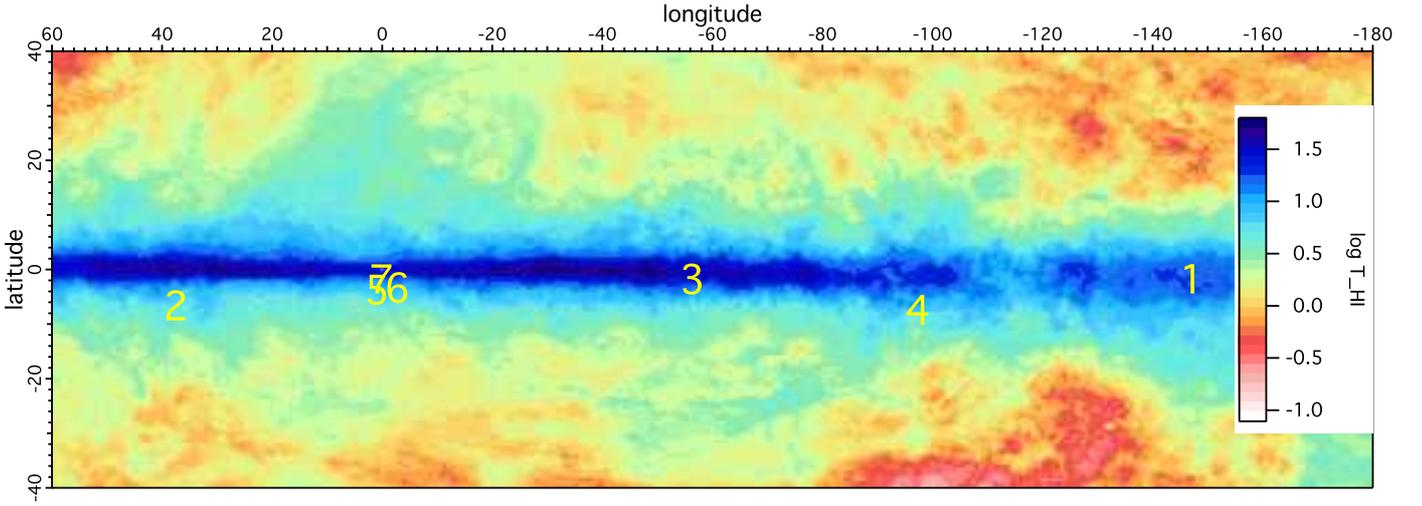} 
\caption{Distribution of the selected fields, numbered as in Table \ref{tabfields}. The distribution is superimposed on the HI 21 cm brightness temperature map (\cite{kal05}) in the radial velocity interval -100 km/s $\leq v_{LSR} \leq$ 100 km/s. The map is in galactic coordinates, centered on $l=-60^{\circ}$.}
\label{stardistribution}%
\end{figure*}

\begin{figure*}[!ht]
\centering
	\includegraphics[width=0.49\textwidth]{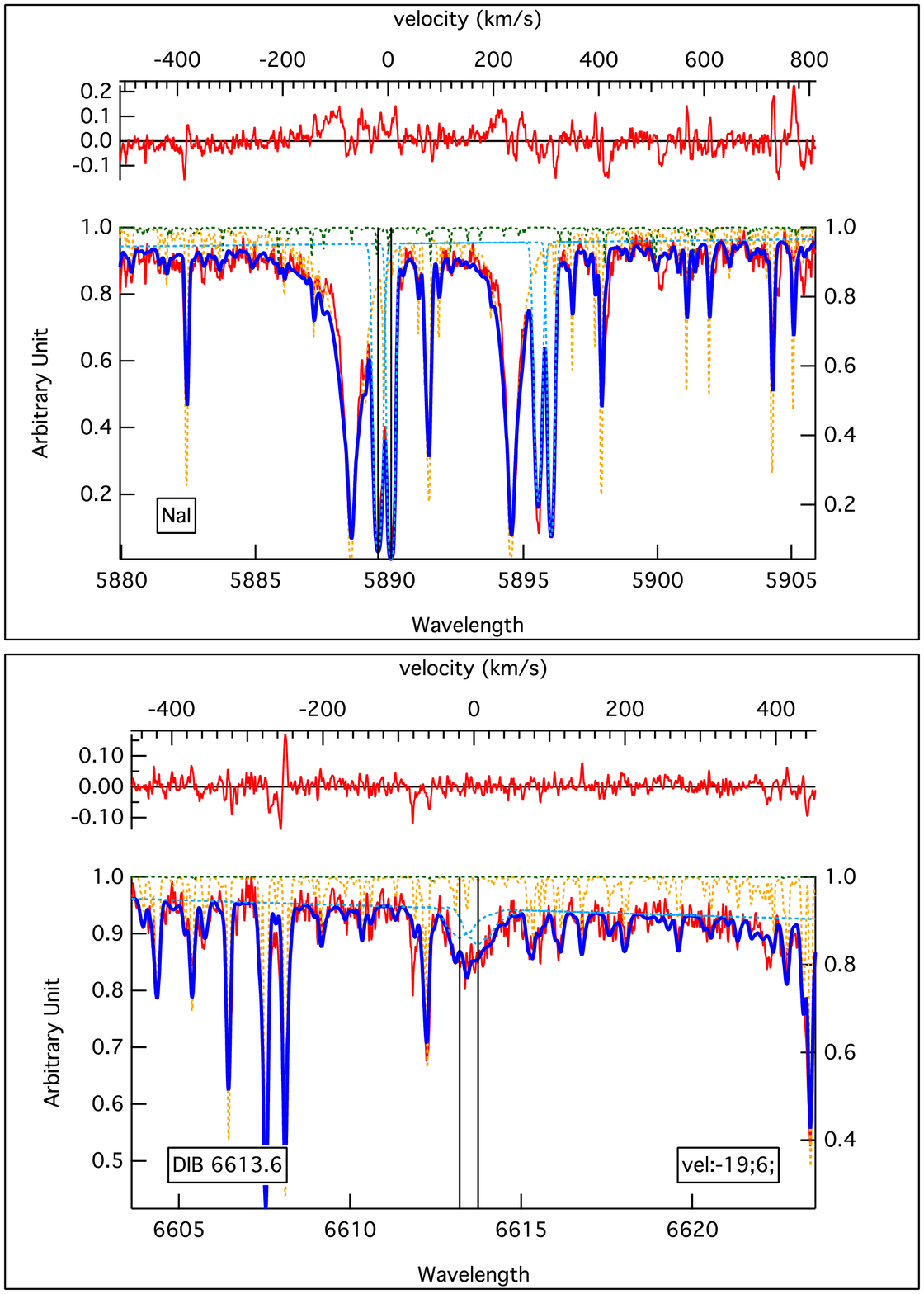}
	\includegraphics[width=0.49\textwidth]{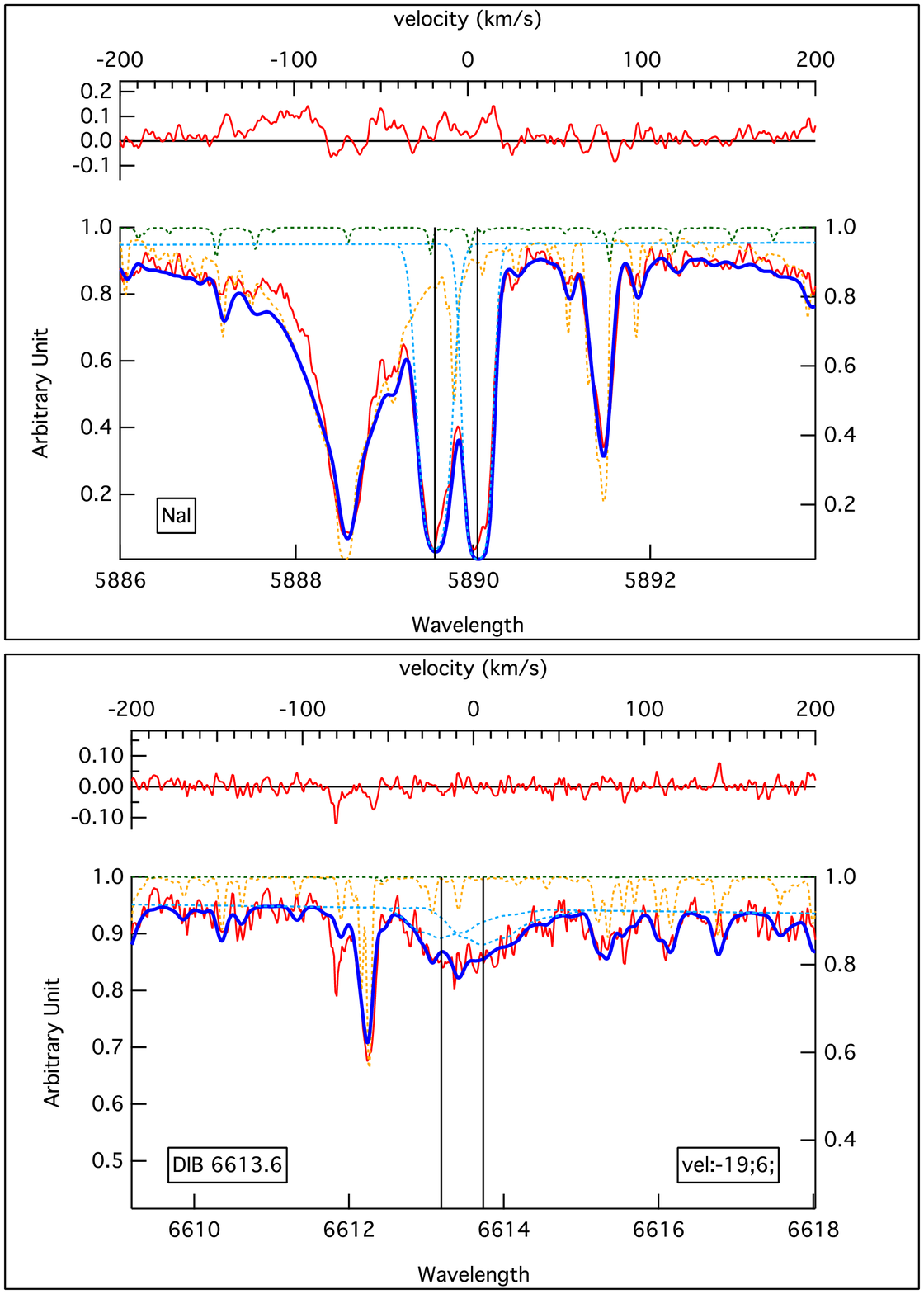}
\caption{Illustration of the NaI doublet/6614 \AA\ DIB global, multi-component analysis, here for the GES star 12574905-6458511 (field 3). The NaI region is shown at top panel and the DIB region at lower panel. The entire fitted spectral interval is shown at left, while the right figure displays enlarged the NaI-D2 region (top) and the DIB region (lower).  
In each figure the red line shows the stellar spectrum (lower plot) and the fitting residuals (upper plot). The dotted lines are the models: stellar (orange), telluric (green), and interstellar components (blue) respectively. The thick blue line is the final adjustment. The radial velocities of the NaI and DIBs components are kept linked (see the black vertical line). Velocities are heliocentric.}
\label{globalanalysis}%
\end{figure*}

\begin{table*}
\caption{Selected Fields with FLAMES Observations. Fields 1 to 5 are from GES. Fields 6 and 7 are part of the ESO program 079.B-0662.}
\begin{center}
\begin{tiny}
\begin{tabular}{|l|c|c|c|c|c|c|c|c|c|c|c|}
\hline 
Field  & GIR.  & UVES   &  $l_c$         & $b_c$         &  D$_{min}$  & D$_{max}$ & A0$_{min}$ &  A0$_{max}$ & Ang. & Setting(s) & Studied    		\\
           & targ.       & targ. &  ($^\circ$)   &  ($^\circ$)  & (kpc)             & (kpc)             & mag              &  mag               &  size ($^\circ$)        &                    & DIBs           \\
\hline 
1 COROT-ANTICENTER &   57  &    7  &  212.9  &    -2.0   &     0.1  &     8.6  &     0.0   &     2.5  & 0.4 & UVES5800,HR15N,HR21 & 6283,6614,8620   \\   
2  COROT CENTER &  105  &    5  &    37.5  &    -7.0   &     0.1  &    16.1  &     0.0   &     2.1  & 0.3 & UVES5800,HR15N,HR21  & 6283,6614   \\   
3 NGC4815 &       &   13  &   303.6  &    -2.1   &     0.9  &     5.0  &     1.0   &     2.5  & 0.1 & UVES5800  &  6283,6614  \\   
4 $\gamma$ Vel &       &   25  &   262.8  &    -7.7   &     0.7  &     2.3  &     0.0   &     1.0  & 0.9  & UVES5800 & 6283,6614 \\   
5 OGLE BUL\_SC45 &       &   12  &     1.0  &    -4.0   &     1.3  &     2.8  &     0.8   &     1.2  & 0.3 & UVES5800  & 6283,6614  \\   \hline
6 OGLE BUL\_SC24(O) &   99  &       &    357.3  &    -3.6   &     0.5  &    10.0  &     1.6   &     3.1  & 0.4 & HR13  & 6283  \\   
7  OGLE BUL\_SC3,4(W) &  106  &       &     0.1  &    -2.1   &     0.7  &     9.6  &     0.7   &     2.7  & 0.4  & HR13  & 6283 \\   
\hline
\end{tabular}
\end{tiny}
\end{center}
\label{tabfields}
\end{table*} 

About 500 diffuse interstellar bands (DIBs) have been detected in the optical domain between 4400 and 8600 {\AA} \citep{hobbs2008,2009ApJ...705...32H,mccall2013}, and their number in the infrared and ultra-violet windows is still growing \citep{joblin90,geballe11}. Identifying the carriers of those irregular features that appear in absorption in stellar spectra is a subject of active research for many years \citep[see reviews by][and references therein]{1995ARA&A..33...19H,sarre2006,2011ApJ...727...33F,camicoxiau}. A lot of effort has been put to extract the most precise information on the DIBs from high resolution, high signal stellar spectra  and derive their various properties, in particular their fine structure and the way they react to the radiation field (see e.g. \citealt{JD94,krelo95,gala00,tuairisg00,cox2005,welty06,2009ApJ...705...32H,vos11}). 
In those spectral studies, DIBs were extracted from hot (early-type) stars because of their smooth, easily fitted continuum. This introduces a limitation on the number of potential target stars that can be used to study DIBs. In the case of nearby stars, it favors highly variable conditions in irradiation and in subsequent DIB carrier destruction or ionization state changes (e.g. \citealt{vos11}). 

Recently, progresses were shown on the extraction of DIBs from cool (late-type) star spectra, in particular a method using synthetic stellar model devised by \citet{chen13}. Such a technique has the advantage of enormously increasing the number of potential targets, probing average conditions in the interstellar medium (ISM) far away from the strong radiation field of UV stars, and simultaneously providing some feedback to improve both the synthetic stellar spectrum and the DIB detection (Monreal-ibero et al., in preparation). Other methods have been applied to cool stars, i.e., using comparisons with unreddened star spectra \citep{2013ApJ...778...86K}, or statistical methods based on principal component analysis (PCA) \citep{zasowski14}.

Independently of the search for their carriers, our goal here is to study how they can be used to trace the ISM at the Galactic scale, both its distribution and its kinematics (see previous works in this direction by \citealt{vanloon,vanloon14,zasowski14}). In particular, DIBs used as an interstellar (IS) tracer may potentially help to build 3D ISM maps by means of inversion methods, similar to the inversion of neutral sodium or extinction data \citep{vergely01,vergely010,2014A&A...561A..91L}.  Thanks to the Gaia mission, launched 19 December 2013, parallax distances should become available for a huge number of Milky Way stars, allowing to build more accurate maps. One of the observational advantages of DIBs over gaseous lines is their spreading over a  wide wavelength interval (from optical to IR), and, more important,  the absence of saturation for distant or particularly opaque sight-lines. 
Another strong advantage over the use of photometric extinction is the derivation of the kinematic information, i.e., the radial velocities of the IS clouds.

All of the individual IS clouds that are present along a line-of-sight (LOS) imprint a specific DIB absorption whose strength and Doppler shift reflect the IS matter content and cloud radial velocity, respectively. This is why measuring DIB equivalent widths in a single-component approach becomes inappropriate when the radial velocity interval that is spanned by all cloud projected motions is not negligible with respect to the DIB spectral width, i.e., it is not a valid technique for narrow DIB and/or distant sight-lines. However, the extraction of multi-component DIBs together with their kinematics has been rarely attempted. \cite{cox2005} used the convolution of a template DIB profile and the multi-component KI absorption profile, while \cite{cordiner08} (resp.  \citealt{cordiner11}) fitted separately the Milky Way and M31 (resp. M33) DIBs using Gaussian profiles. Here, we present improved fitting methods allowing for multi-component of DIBs.  The methods are fully automated. Automated here means that no intervention by the user is needed during the series of fitting that are launched in a unique run for a large number of spectra. More precisely, no spectral interval selection for continuum  fitting is needed, and there is a total absence of manual "guesses" (most profile-fitting methods are only partly automated and require those "manual" steps). Each component has a pre-determined shape derived from high resolution spectra of hot nearby stars. The methods are suitable for any type of stars as long as their stellar parameters have been determined and their synthetic spectra can be computed. 


We have applied these new fitting techniques to a series of spectra of cool target stars for which stellar atmospheric parameters and estimated distances have been determined spectroscopically.  
Part of the data are from the Gaia-ESO Spectroscopic survey  (GES) \citep{2012Msngr.147...25G}, a public spectroscopic survey that started in 2011 and that aims at recording VLT/FLAMES spectra of $\sim$ 100000 stars in our Galaxy down to magnitude 19, systematically covering all the major
components of the Milky Way and the selected open clusters. This survey will provide 
a wealth of precise radial velocity and abundance determinations. The other data is part of an earlier program devoted to the study of the inner disk \citep{hill12,hill14}. Deducing properties of the ISM is a \textit{by-product} of these stellar-oriented programs. 

Seven FLAMES fields were selected for being widely distributed in galactic longitudes, to probe very different interstellar cloud complexes, and close to the Plane, to ensure significant absorptions. They were chosen totally independently of the primary objectives (i.e. open cluster studies, bulge star properties, etc.., and of the target star properties themselves). We also gave priority to fields with targets widely distributed in distance.
Our goal is (i) to test our interstellar absorption fitting methods, (ii) to study the variation of the DIBs as a function of the distance along the LOS and show the potentiality of the DIBs for 3D mapping purposes, and (iii) to study the DIB-extinction relationship in different regions of the Milky Way (MW) disk. 

Section 2 presents the data and some general properties of the selected DIBs. Section 3 describes the spectral analysis method for multi-component DIB extraction and illustrates its application. Section 4 describes the results and the observed DIB properties. In this section we compare the DIB equivalent widths with the estimated extinctions and draw LOS profiles of DIBs in the various directions. Section 5 discusses future improvements and the mapping potentialities.

\section{Data and choice of DIBs}

Of the seven fields, five fields are GES data.
We complemented the GES data with previously recorded spectra from two fields towards the bulge. 
Along with one of the GES LOS, this allows comparisons between DIBs in directions that differ by a few degrees.  Overall we tried to probe 
a variety of cases to test our methods. All of the selected spectra are characterized by a good signal-to-noise (S/N) ratio, S/N $\gtrsim$ 50, which ensures good results.

Figure \ref{stardistribution} shows the distribution of the fields in the sky, superimposed on a HI 21cm emission map.  The projections to the Plane are also shown in Fig \ref{gxmap}. All targets were observed with the FLAMES multi-object spectrograph at the VLT-UT2. We used both GIRAFFE  ($R \simeq 17000$)  and UVES ($R \simeq 47000$) observations \citep[see][for UVES]{2000SPIE.4008..534D}. The UVES spectra cover  the 5822 to 6831 \AA\ spectral range which contains the \textit{classical} NaI (D2-D1 5889.9-5895.9 \AA) IS lines as well as some rather strong DIBs, such as the 6283.8 (hereafter called 6283) and 6613.6 (6614) \AA\ bands. Depending on the observed field, the GIRAFFE observations were made with the H665 (HR15N) setting (spectral range 6444-6816 \AA) which allows study of the 6614 \AA\  DIB at a lower resolution than UVES, and with the H875 (HR21) setting (spectral range 8475-8982 \AA) which includes the 8620.4 (8620) \AA\ DIB (informally known as the \textit{Gaia} DIB, since it is contained in the spectral interval of the Radial Velocity Spectrometer (RVS) on board the satellite). The additional inner disk bulge data was observed with GIRAFFE H13 setting (spectral range 6170-6309 \AA). 
The GES UVES and GIRAFFE reduced spectra are issued from the dedicated pipeline \citep{sacco}, while the two OGLE field spectra were reduced by means of using our dedicated  GIRAFFE  tool based on the ESO pipeline. 
  
 
Table \ref{tabfields} lists the selected fields, the number of targets in each field, the field center coordinates, the observing modes, and the whole range of estimated stellar distances and extinctions (see the next section). The full list of target stars along with their coordinates, estimated extinction and distances can be found in the online Appendix. There are 429 target stars, from which about half (224) have been observed as part of GES. A majority of those GES target stars are within the GES-CoRoT (COnvection ROtation et Transits plan\'etaires) fields (172 stars).



We focus on the 6614  and 6283 \AA\ DIBs that are strong enough to ensure a detection in most targets. When recorded, we also analyzed the shallower 8620 \AA\ DIB. 
The 6614  \AA\  DIB  
is a widely studied, strong and narrow DIB and has a good correlation with E(B-V) \citep[see][etc]{sonnentrucker97,2011ApJ...727...33F,vos11,2013A&A...555A..25P,2013ApJ...774...72K}. 
The broader 6283 \AA\  DIB is 
is a strong, broad DIB that was also widely studied and is known for being significantly influenced by the radiation field (\citealt{vos11}). The 8620 \AA\ DIB 
is rather weak band that has been recently studied as part of the RAVE spectroscopic Survey \citep[see][]{2008A&A...488..969M, 2013ApJ...778...86K} and is of particular interest in the frame of Gaia. It seems to be quite well correlated with the reddening, although the number of studies is still limited.

\section{Data analysis}

\subsection{Description of the fitting method}

The principles of the fitting method are essentially  the same as in \cite{chen13}, 
the main difference being that we allow here for multi-component DIBs, and subsequently extract kinematic information. 
As the length of LOS increases, differences in cloud radial velocities may become comparable or larger than the DIB width, making the use of a multi-component fit necessary.
We model the observed spectrum as the product of a synthetic stellar spectrum ($S_{\lambda}$), a synthetic telluric transmission ($T_{\lambda}$), and a DIB model that is itself the product of several DIB profiles, each one representing one absorbing cloud complex. When the telluric absorption is very weak or negligible, $T_\lambda \simeq 1$. Finally, to take into account the local slope of the unnormalized spectrum, we allow for a continuum that is simply represented by a linear polynomial with A and B as the coefficients. This appears to be sufficient for our limited wavelength interval around each DIB.  The model spectrum ($M$) can be therefore written as,
\begin{eqnarray}
M(_{\lambda})= S_{\lambda} [V_{star}] \hspace{0.2cm}\times\hspace{0.2cm} T_{\lambda} [V_{tell}]^{\alpha_{tell}}  \hspace{0.1cm}\times\nonumber \\
 \hspace{0.4cm}  \Pi^{i}(DIB^{i}_{\lambda} \hspace{0.05cm}[vel^{i}]^{\hspace{0.05cm}\alpha^{i}}) 
  \hspace{0.2cm} \times\hspace{0.2cm} ([A]+[B]\times\lambda) \hspace{0.2cm}.
\end{eqnarray}
$V_{star}$ is the stellar radial velocity, $V_{tell}$ is the Earth's motion, and $vel^i$ is the interstellar cloud radial velocity. These various terms are detailed below, as well as the coefficients $\alpha_{tell}$ and $\alpha_{i}$.

The computation of the stellar model $S_{\lambda}$ requires the preliminary knowledge of the stellar parameters.  For each of our target stars, the effective temperature, gravity, metallicity, and micro-turbulence have been previously determined: (i) for the GES targets we use the 
 stellar parameters jointly determined by the GES team members \citep{smiljanic14,recioblanco14,Lanz2014}; (ii) for the additional archival data, see \citet{hill12}.  Based on the stellar parameters, a synthetic stellar model was computed for each target star using an ATLAS 9 model atmosphere and the SYNTHE suite \citep{2005MSAIS...8...14K,sbo2004,sbo2005}. In the case of GES targets, this may yield a synthetic spectrum which is not exactly the same as the one of the synthetic spectral library used in GES. Similarly, inner disk spectra may be slightly different from those used in the first analysis. 
However, in both cases the differences should be too small to influence the determinations of the DIBs, see section 5. 

The synthetic telluric transmissions $T_{\lambda}$ were computed by means of the LBLRTM code (Line-By-Line Radiative Transfer Model, \citealt{lblrtm05}), using the molecular database HITRAN (HIgh-resolution TRANsmission molecular absorption \citep{hitran2008}. This telluric transmission model is available online in TAPAS web-based service \citep{tapas}. Telluric lines are strong in the 6283 \AA\ spectral region and  negligible for the  6614 \AA\ band. We make use of the same telluric models for the derivation of the fitting of neutral sodium lines. The coefficient $\alpha_{tell}$ is proportional to the optical depth of the telluric lines.

The models for the 6614 and 6283 \AA\  bands are empirical profiles that have been previously determined from high signal to noise spectra of nearby stars \citep{2013A&A...555A..25P}.
Since the laboratory wavelengths for the DIBs are currently unknown and their profiles are irregular, the choice of rest wavelengths that correspond to a null Doppler shift of the absorbing matter is somewhat arbitrary. Throughout this work, we use, for these first two DIBs, the wavelength values derived by \cite{hobbs2008} who cross-calibrated the DIB profiles and interstellar KI absorption lines. We assumed that the rest wavelength corresponds to the deepest point in the profile. Because our model profiles may slightly differ from the  \cite{hobbs2008} profiles, a small offset may exist between the rest wavelengths, of the order of few km/s, that we neglect here. On the other hand, it is well established that the 6614 \AA\  DIB  has substructures, and that these substructures may slightly vary from one LOS to the other \citep{2002A&A...384..215G}. This results in small changes of the overall profile. In our case, the GIRAFFE and UVES spectral resolutions do not allow these subtle changes to be distinguished. We ignore the profile variability to simplify the modeling. For at least the 6614 \AA\ DIB, it has been shown that in very rare, extreme conditions for the radiation field, the DIB profile may evolve and be characterized by a redward wing \citep{oka13}. We neglect this possibility here, a reasonable assumption since our  LOS do not target 
particular strong infrared sources. 
The model for  the 8620 \AA\  DIB is also an empirical model, obtained by averaging DIB profiles from several spectra based on the \cite{chen13} data analysis. For this band the rest wavelength is chosen to be the one defined by \cite{2008A&A...488..969M}. 
The three empirical DIB profiles are defined over the $\lambda\lambda$ 6609-6619 \AA, 6263-6303 \AA, and 8612-8628 \AA\ intervals respectively. Finally, $\alpha_{i}$ is an adjustable coefficient that is the ratio between the optical depth of the absorber that produces the DIB and the optical depth of reference.

The fitting procedure adjusts to the data the convolution of the above product by the instrumental function, here represented by a Gaussian ($G$). During the adjustment of the composite stellar-DIB-telluric model, we allow Doppler shifting of the stellar model by a free quantity $V_{star}$ to take into account the stellar radial velocity, of the telluric transmission model by a free quantity $V_{tell}$ to take into account the Earth's motion, and of the DIB profile $i$ by a  radial velocity $vel^{i}$  to take into account the ISM kinematics. We could evidently use the star radial velocity that comes out from the stellar spectrum analysis and is derived over a much wider wavelength range, and we could also make use of the telluric information linked to the observing conditions.  However, 
a cross-correlation operation has been actually integrated in our code to make a first estimate of these offset values which is convenient for handling any spectroscopic data
, and allow for their fine tuning during the adjustment. 
Our derived values actually conform to the expected ones. We allow for changes of the $\alpha_{tell}$ parameter and $\alpha^{i}$ to adjust the telluric lines and DIB strength, respectively. 

The DIB equivalent width (EW) is derived in two different ways: 
(i) by using the best fit DIB strength $\alpha^{i}$ and the equivalent width of the DIB model, which provides a first result we refer to as the fitted EW ($EW_f$),  or (ii) by measuring the true area of the absorption band with respect to the continuum, which provides a second result that is independent of the DIB model and we name the continuum-integrated EW (EW$_{ci})$. The $EW_{ci}$ is obtained after subtraction of the other components (stellar and telluric lines) in the normalized spectrum. 
In the multiple component case, the EW for each absorbing cloud can be independently derived using the $EW_f$ method. The total of intervening matter corresponds to the sum of the fitted EWs from each DIB component. In contrast, the $EW_{ci}$ method does not detect the individual components, but only measure the total absorption.   
The spectral interval used for the computation of the DIB EW is the same as the one of \cite{2011ApJ...727...33F}, or in the case of 8620 \AA\ DIB is taken from - 7 to +7 \AA\ from the DIB center.

In principle, sky emission lines disappear after background subtraction, however there are potential residuals. The spectral ranges we consider here for the DIB extraction are free of
strong sky emission lines, e.g. OI at 6300 \AA\ does not  overlap with the 6283 \AA\ DIB. There is an exception in the case of the red wing of the 8620 \AA\ DIB where emission line residuals may influence the DIB fitting (see next sections).  Similarly, there may be a presence of features within strong stellar lines which are not accounted-for by stellar atmosphere models, e.g. circum-stellar H alpha 
emissions or interstellar permitted and forbidden emissions, but they do not overlap with our selected DIBs.


\subsection{Fitting strategies and examples of adjustments}

For all multi-component adjustments, it is necessary to start with initial parameters that are as close as possible to the actual solutions, in order to avoid secondary minima and in order to converge more rapidly toward the final solution. Here, the initial guesses for the number of required velocity components, their radial velocities and strengths come either from interstellar NaI lines as in the case of UVES spectra, or, in the absence of any absorption line, from a simplified decomposition of the HI emission spectrum, taken from the spectral HI cube in the direction of the target star as in the case of GIRAFFE spectra. Prior to the use of those guesses we performed profile-fitting tests without any such initial parameters, and compared with the subsequent results. We did not find any negative influence of the guesses such as biases towards a non realistic solution, instead we always found the expected positive effects of fast convergence towards the primary minimum.

\subsubsection{Use of the NaI absorption lines}

In the case of the NaI  lines, they are not only used as sources of the first \textit{guesses} of the cloud parameters, but they also enter in the global analysis of lines and DIBs, which means that they are simultaneously measured together with the DIB components. Their radial velocities are linked to remain identical throughout the adjustment, component by component. Such a method is justified by the fact that any  NaI line must have a (strong or weak) DIB counterpart. From the previous observations, we know that all of the detected DIBs were found to be associated with strong neutral sodium lines. There may be a small Doppler shift between the DIB and interstellar NaI line center due to the preferential presence of the DIBs carriers in a particular phase, e.g. at the cloud periphery or in the core. However, those shifts remain small compared with the DIB widths and we will neglect this effect. On the other hand, such a global fitting method is particularly tractable here because the determination of the initial guesses for the parameters can be quite precise, especially if the interstellar lines used are not saturated. 

\begin{figure}[h]
\centering
	\includegraphics[trim=0cm 0cm 0cm 0cm, clip=true,width=0.9\linewidth]{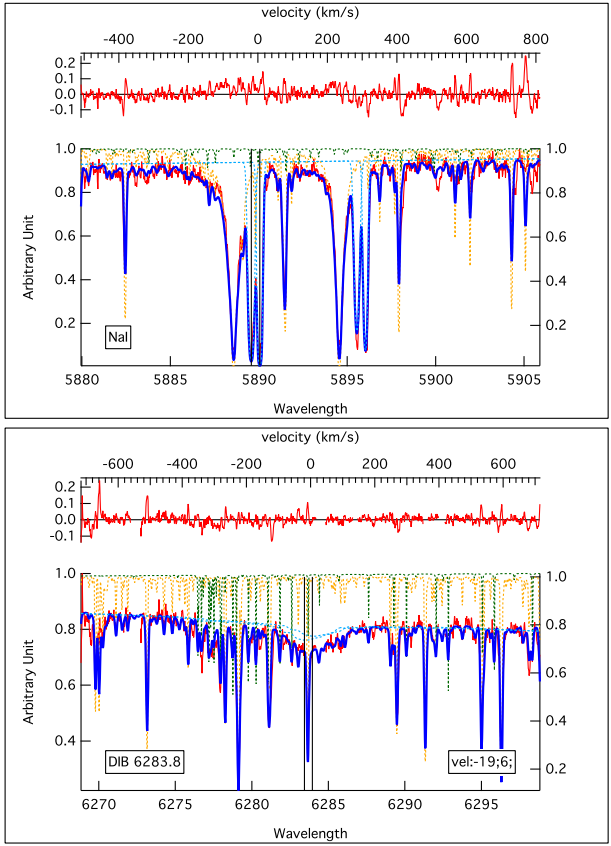}
\caption{Same as Fig. \ref{globalanalysis} (except for the enlarged figure) for the 6283 \AA\ DIB.}
\label{globalanalysis2}%
\end{figure}

The automated global analysis procedure is developed in the frame of the \cite{igor} software and environment which allows to fit multiple data sets simultaneously while linking some of their parameters. Initial guess values for the radial velocities of the interstellar NaI components are preliminarily determined from the observed spectrum on the basis of the main absorption peaks. 
The sodium lines are modeled by Voigt profiles with three free parameters: opacity, radial velocity, and apparent temperature. In normal, realistic profile fitting of NaI lines, the apparent temperature (combination of thermal broadening and turbulence) is constrained to be $T < 10000$K, since NaI is negligible in warmer gas. However, here we are interested only in the first order kinematics, and neither the actual number of clouds nor the NaI columns need to be known in details.  
This is why, in order to avoid having too many interstellar components, we extend the line broadening and allow for a significantly higher apparent temperature ($T < 100000$K). In turn, we list EWs only, and omit NaI column densities that are too imprecise. Figure \ref{globalanalysis}  shows an illustration of the global analysis of the NaI- D2/D1 lines and the 6614 \AA\ DIB. In all cases the fitting results reveal a good agreement between the DIB/NaI radial velocities and the main HI 21 cm velocities. Figure \ref{globalanalysis2} is a similar illustration for the 6283 \AA\ DIB. 

\subsubsection{Use of the HI 21 cm emission profiles}

In the second case, i.e. when no NaI lines are available and instead HI emission spectra \citep{kal05} are used, the fitting scheme is different. Since the HI emission spectrum represents the totality of the IS clouds, both in front of and beyond the target star, a global analysis based on all main HI components is inappropriate. 
The HI emission spectrum is therefore used to construct a table of velocity guesses 
 ($v_{r_{HI}}$) 
and provide upper and lower limits to the velocity range. Then, the DIBs are fitted independently of the actual HI measurement, using a hierarchical sequence of velocity prior values described below.  Another significant difference in this second case is that the initial values of the cloud Doppler shifts are much less precise than in the case of sodium lines, and the cloud velocity profiles strongly overlap (for the same gas temperature the Doppler width is about 5 times wider than for sodium). Finally, the HI map has a spatial resolution of $\sim 0.6^\circ$, larger than the FLAMES field-of-view. 
Still, 
the emission spectra give an appropriate starting point for the fitting and initial parameters of the interstellar cloud components. 

\begin{figure}[h!]
\centering
	\includegraphics[width=0.9\linewidth]{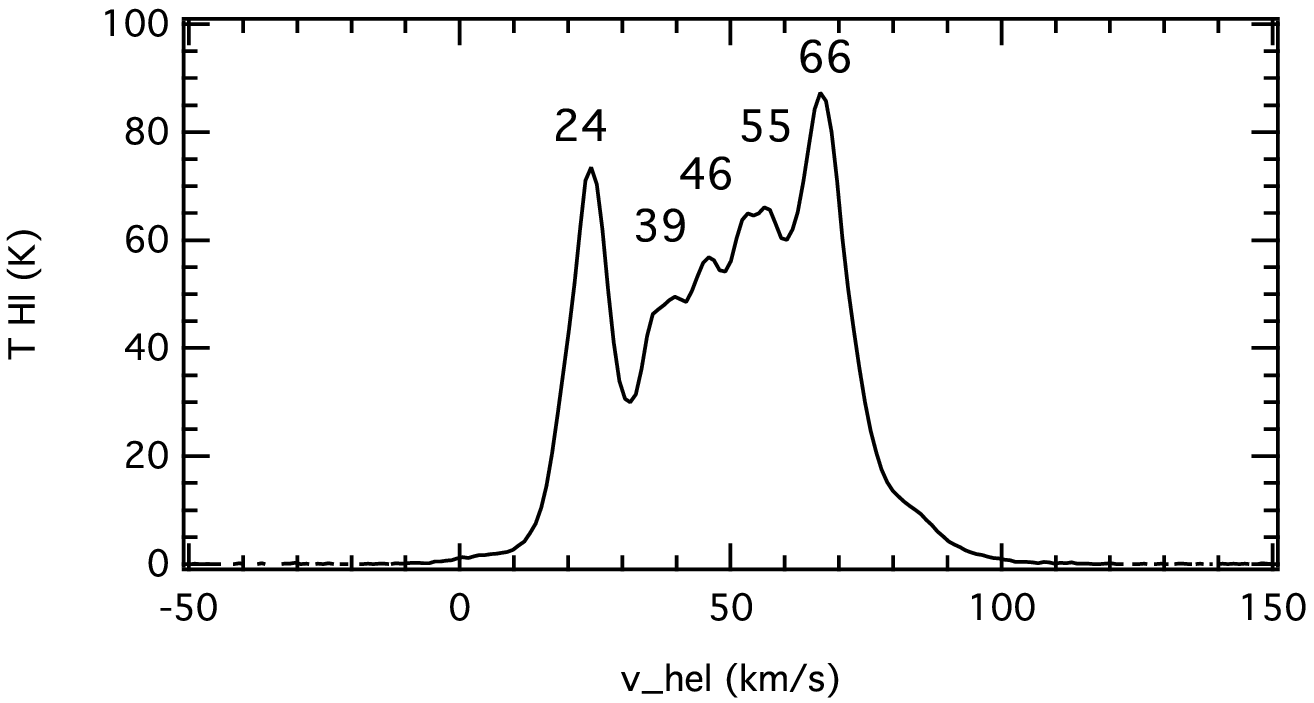}\\ 
	\includegraphics[width=0.99\linewidth]{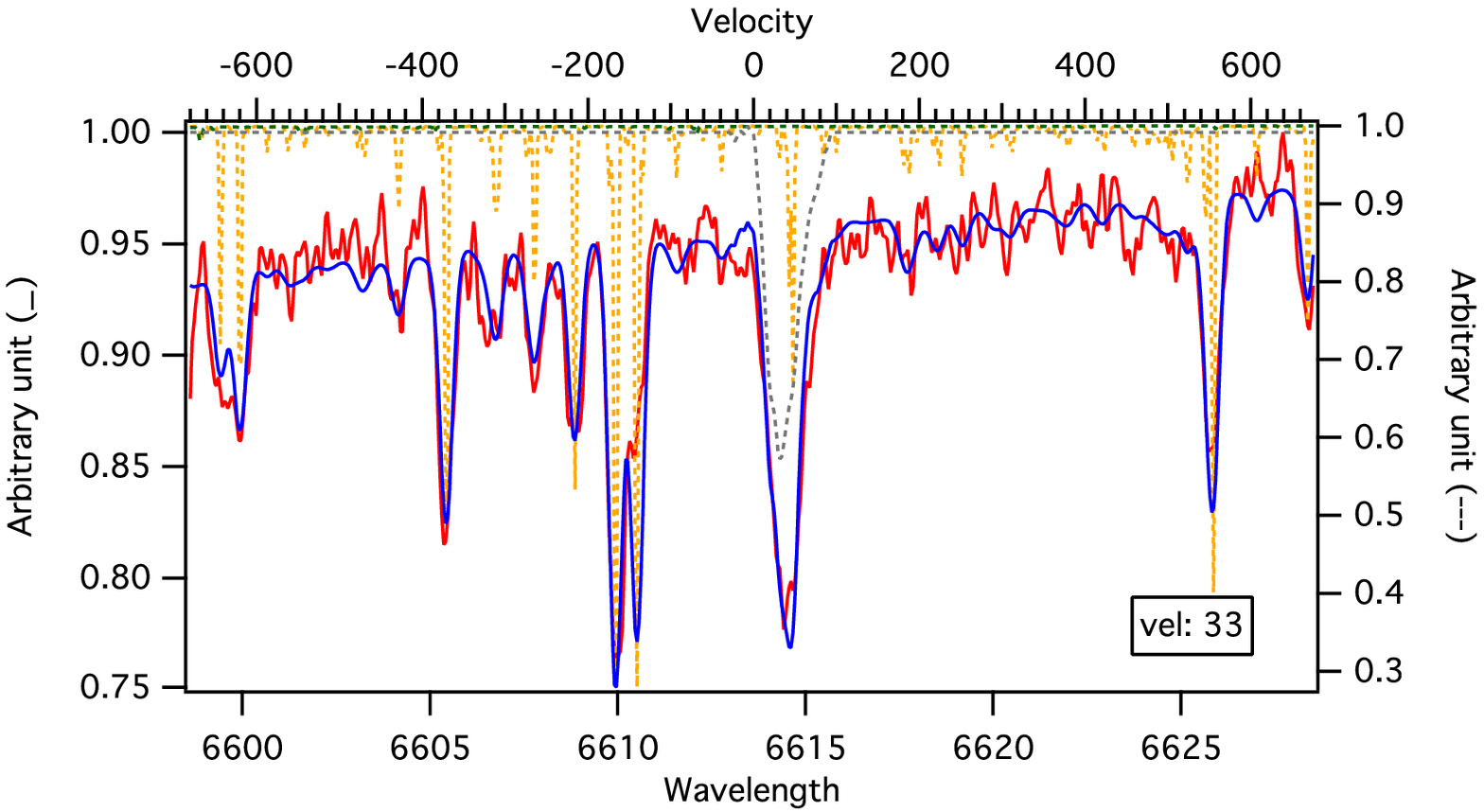}
	\includegraphics[width=0.99\linewidth]{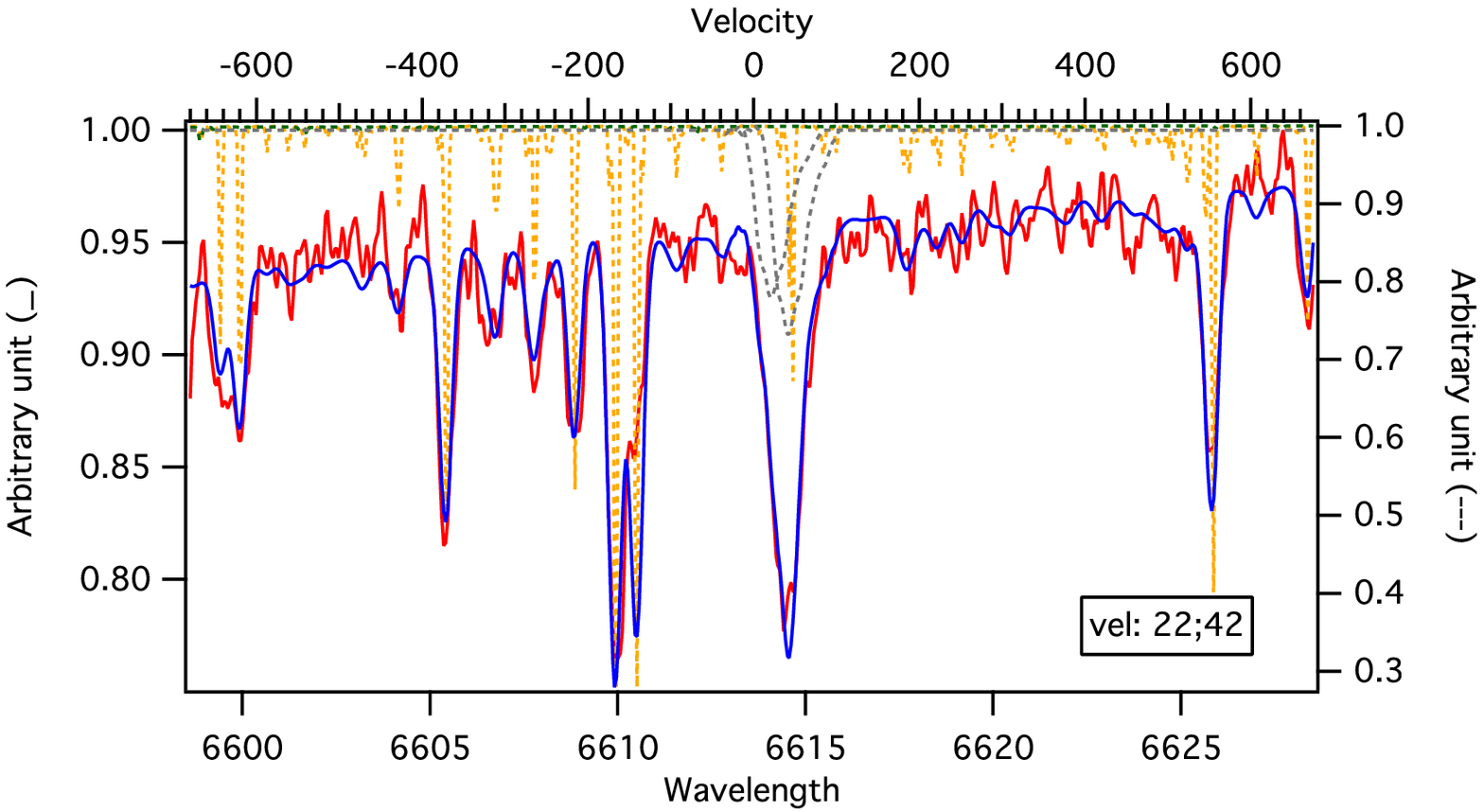}
\caption{Illustration of the multi-component 6614 \AA\ DIB fitting: GIRAFFE field 1, GES target 06441034-0048254. The red line shows the stellar spectrum. The dotted lines are the models: stellar (orange), telluric (green), empirical DIBs (grey). The thick blue line is the optimal model adjustment. The initial guesses for  the DIB velocity centroids are 24, 40, 50 km/s and based on the HI spectrum in the same direction (see top plot). 
\textit{Middle:}  an example of a preliminary adjustment with a unique DIB component. The DIB velocity is found to be $\sim$ 33 km/s. The large difference from the initial guess (24 km/s) demonstrates the need for the introduction of a second cloud component. \textit{Bottom:} an example of a subsequent adjustment with two DIB components. The two fitted velocities are now close to the first two HI emission peaks.}
\label{multiDIB}%
\end{figure}

However, the 6614 and 8620 \AA\  DIBs have very different widths and only the 6614 \AA\ DIB is narrow enough that multi-components with velocity differences on the order of 10 km/s or more can be distinguished in an automated way. Figure \ref{multiDIB} shows an example of such a fitting of this DIB based on the HI initial guesses. The first adjustment involves a single component $v_{r_{HI}}$ and uses as a guess the smallest absolute value of the HI velocities, that in all cases corresponds to local gas. When the single-component velocity derived from the fit is significantly different from $v_{r_{HI}}$, the second velocity component from the $v_{r_{HI}}$ table is included and a fit with those two prior 
values is performed, and so on. Using two components gives a significantly better fit, see the red part of the DIB.

\begin{figure}[h!]
\centering
	\includegraphics[width=\linewidth]
{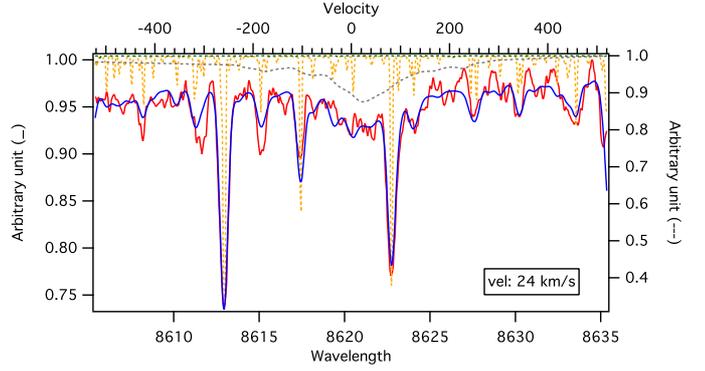}
	\caption{Same as Fig. \ref{multiDIB} (middle), but for 8620 \AA\ (GES target 06441034-0048254). We remark sky residuals in the red wing of DIB. For this broad DIB a single velocity component is used (see text).} 
\label{multiDIB8620}%
\end{figure}

The very broad 8620 \AA\ band does not react with enough sensitivity to changes on the order of 10-20 km/s for  guiding the fit to multi-velocity solutions, at least for our present dataset. Moreover, many spectra are contaminated 
by sky emission residuals which make the fitting even more difficult. For those reasons and after several negative tests we have chosen to keep the mono-cloud procedure
, i.e. we consider only the first step (see Fig. \ref{multiDIB8620}) and the prior is the velocity that corresponds to the smallest absolute value (the local value). Still, the derivation of the DIB EW is made with a rather good precision, as tests made with one or more components have shown, again due to the large width of this absorption band. Exceptionally,  we use the velocity results from the 6614 \AA\ DIB fitting as the initial guesses of the 8620 \AA\ DIB fitting to avoid the artificial effects of the sky emission contamination.

\subsection{Derivation of the DIB equivalent width and error estimate}

As previously discussed, the DIB EW can be derived using two different ways: the EW$_f$ and the EW$_{ci}$. The 
results and figures that are presented in this article all correspond to the first method. As already said it allows for the distribution into separate components, but it has also the additional advantage of being less influenced by the potential errors in the computed stellar lines. 
The reason why both EWs are computed in each case is that their comparisons  acts as a flag for the quality of the fit and reveals bad quality spectra. For all data we present the two EWs are found to agree within the observational and model uncertainties. 

The errors on the EW have three distinct sources: errors on the stellar continuum determination, statistical noise, and errors on the stellar model: 
$\sigma^2=\sigma_{cont}^2+\sigma_{S/N}^2+\sigma_{stellar}^2$.
The error on the stellar continuum placement is mainly linked to the statistical noise and both errors are estimated in a joined manner. In order to obtain a first, global estimate of those combined errors, we performed a preliminary study 
that is a series of simulations with varying random noise. 
For each simulation, we fitted the DIB and then compared all resulting EWs. For a random noise representative of the typical S/N of the spectra	(S/N$\simeq$100), 
we obtained a typical relative error of about 5\% on the EW (more specifically a deviation of $\sim$ 5 m\AA\ when the EW is 100 m\AA). This gives an 
estimate of the contribution of the first two errors. 
Regarding the third error linked to the stellar model, we already know that
the data-model residuals are larger than average for some specific stellar lines and depend mainly on the stellar effective temperature and metallicity \citep[see][]{chen13}. 
Figures  \ref{residual} and Fig. \ref{residual2} in the Appendix show the stacked residual of the DIB fitting for the 6614 \AA\ and 8620 \AA\ bands for $\sim 160$ GIRAFFE stars, and examples of dependence on the star effective temperature. To study the order of magnitude of the contribution of the stellar model to the error, we extracted %
all of the residuals, and estimated their maximum level at the center of the DIB. This corresponds to the most contaminated cases for which the stellar line falls close to the DIB center. Then we performed again a random noise simulation with this new variance instead of the measurement noise, and we obtained a new error estimation on the order of 13 \% and 15\% respectively for the two DIBs. We estimate that this gives us a realistic estimate of the maximum total error from the three sources.
Although our final estimate for individual spectra will be based on another method, described below, 
this range for the errors linked to the signal and the model illustrates the gain in precision we can expect in future by improving the stellar model.

Applying the above method to each of individual targets would be too time consuming. Instead, we use a different approximation. For the first two errors, we use the following formulation, $\sigma_{S/N+cont}=\sigma_{S/N} \times \frac{\Delta\lambda}{\sqrt(N)}$; $\Delta \lambda$ is width of DIB and $N$ is number of points/pixels covering this width. The signal to noise ratio is estimated for each spectrum based on a linear fit in a clean area. Secondly, to obtain the third error, $\sigma_{stellar}$, we performed two consecutive fits without, then with 
masking of the strong stellar lines that fall in the DIB interval. The number of masked lines depends on the DIB and on the stellar radial velocity (it varies between one and three lines). 
The difference between the two calculated EWs (when stellar lines are not masked and when they are masked) gives us an estimate of $\sigma_{stellar}$, $\sigma_{stellar}=\Delta EW_f = EW_f - EW_{f masked}$. Finally the total error is 
$\sigma^2 = \sigma_{S/N+cont}^2 + \sigma_{stellar}^2 $. This method gives errors that are in agreement with those from the preliminary study described above.

For the 8620 \AA\ DIB, we have an additional complication because the right wing of the DIB region is sometime contaminated by sky emission residual $\sigma_{sky}$. Correcting for this emission is beyond the scope of this work, and estimating its effects can not be done in the same way as for the two other DIBs 
because the contamination results in a "bumpy" feature which changes the absorption shape and produces a non realistic runaway shift from the true DIB radial velocity.
Instead, we calculated the error as the sum of four terms: $\sigma^2=\sigma_{cont}^2+\sigma_{S/N}^2+\sigma_{stellar}^2+\sigma_{sky}^2$. The term $\sigma_{sky}$ is obtained by calculating the variance in the region of sky contamination and multiplying by the full width half maximum (FWHM) of the DIB profile. We plan to incorporate in future the pixel-by-pixel estimated uncertainties provided by the pipeline.

\subsection{Distance and reddening estimates}


To estimate the distance and extinction of the GES data, we used the 2D Bayesian method described in \cite{Babusiaux14} \citep[see also][]{2010MNRAS.407..339B}. All our targets have 2MASS NIR photometry \citep{Cutri03} as well as V magnitude from different sources: OGLE-II photometry \citet{Udalski02} for the bulge directions, \citet{Deleuil09} for the CoRoT fields, and \citet{Bragaglia14} for the open clusters. We will use here the V-K colour which is more sensitive to the extinction than the J-K colour used in \citet{Babusiaux14}.

We used the \cite{Bressan12} isochrones (Parsec 1.1) with a step of 0.05 in log(Age) between [6.6, 10.13] and a step of 0.05 dex in [M/H] between [$-2.15$, 0.5]. Each isochrone point $i$, corresponding to a metallicity [M/H]$_i$, age $\tau_i$ and mass $\mathcal{M}_i$, has a weight associated to it $P(i)$ according to the Initial Mass Function (IMF) $\xi(\mathcal{M})$ and Star Formation Rate (SFR) $\psi (\tau)$. We used here the \cite{Chabrier01} lognormal IMF (integrated over the mass interval between isochrone points) and a constant SFR (considering that we have a grid sampled in logAge this means that the SFR associated weight is proportional to the age), and we did not introduce any age-metallicity correlation.

We computed the probability of a star with the observed parameters $\tilde{O}$ ($\widetilde{\Teff}$,$\widetilde{\logg}$,$\widetilde{\FeH}$,$\tilde{V}$,$\tilde{K}$) to have the physical parameters of the isochrone point $i$ ($\Teff_i$,$\logg_i$,$\FeH_i$, $\tau_i$, $\mathcal{M}_i$, $V_i^0$, $K_i^0$),
\begin{equation}
P(i|\tilde{O}) \propto P(\tilde{O}|i) P(i). 
\end{equation}
To compute $P(\tilde{O}|i)$, we assume Gaussian ($\mathcal{N}$) observational errors $\epsilon_O$ on the atmospheric parameters and the magnitudes. 
Assuming a distance $d$ and an extinction $A_{0}$ for the isochrone point $i$, we have
\begin{equation}
 P(\tilde{O}|i,d,A_{0}) \propto \prod_O \mathcal{N}(\widetilde{O}-O_i,\epsilon_O).
\end{equation}
However the atmospheric parameters derived from spectroscopy ($\widetilde{\Teff}$,$\widetilde{\logg}$,$\widetilde{\FeH}$) are not independent. For the inner disc fields we derived correlation coefficients that we applied in the above equation using a multivariate normal distribution. For the GES UVES parameters, the GES Consortium provides the individual node values, so instead of using only the recommended value we use all nodes individual values (in general around 5 nodes provides parameters for the same star), which mimick the correlation we want to introduce on a star by star basis. For the GES GIRAFFE parameters we have no information about the correlations available.


The apparent magnitude $m_i$ derived from the isochrone $i$ is a function of the absolute magnitude $M_i^0$, the extinction $A_m$, and the distance $d$:
\begin{equation}
m_i = M_i^0 + 5 \log d -5 + A_m.
\label{eq:Pogson}
\end{equation}
We therefore derived $P(\tilde{O}|i,d,A_{0})$ for a very thin 2-D grid of distances $d$ and extinction $A_{0}$. $A_{0}$ is the absorption at 5500 \AA\, and is roughly equivalent to $A_V$ (e.g. \citealt{CBJ11}).
To derive the extinction in the different photometric bands $A_m$, we used the extinction law  $E_\lambda = 10^{-0.4 k_\lambda}$ of \cite{FitzpatrickMassa07}. We used a typical red clump SED $F_\lambda^0$ from \cite{CastelliKurucz03} ATLAS9 models.
With $T_\lambda$ the photometric total instrumental transmission we have
\begin{equation}
A_m  = -2.5 \log_{10}\left({\int F_\lambda T_\lambda E_\lambda^{A_{0}} d\lambda \over \int F_\lambda T_\lambda d\lambda}\right).
\end{equation}
To take the non-linearity of the above equation into account, we used a discrete table of $A_m$ as a function of $A_{0}$. No prior on the distance nor extinction is added.



What we seek is the distance probability $P(d,A_{0}|\tilde{O})$, which we obtain by marginalization over the isochrone points:
\begin{equation}
P(d,A_{0}|\tilde{O}) \propto \sum_i P(\tilde{O}|i,d,A_{0}) P(i).
\end{equation}

Marginalization over the extinction leads to $P(d|\tilde{O})$ and marginalization over the distance leads to $P(A_0|\tilde{O})$. The resulting distance and extinction estimates used hereafter corresponds to the mode of the distribution and the errors corresponds to the 68\% highest Bayesian confidence interval (or highest density interval, HDI).

\section{Results}

The first subsection discusses the measurements of the two CoRoT fields, whereas the second subsection discusses the measurements of the five other fields that have less targets.

\subsection{CoRoT Fields}


DIBs in these sight-lines were derived following the fitting strategy described above. 
All of the measured EWs, uncertainties and NaI/DIB velocities are listed in the Appendix.

\begin{figure}[h!]
\centering
\includegraphics[width=\linewidth]{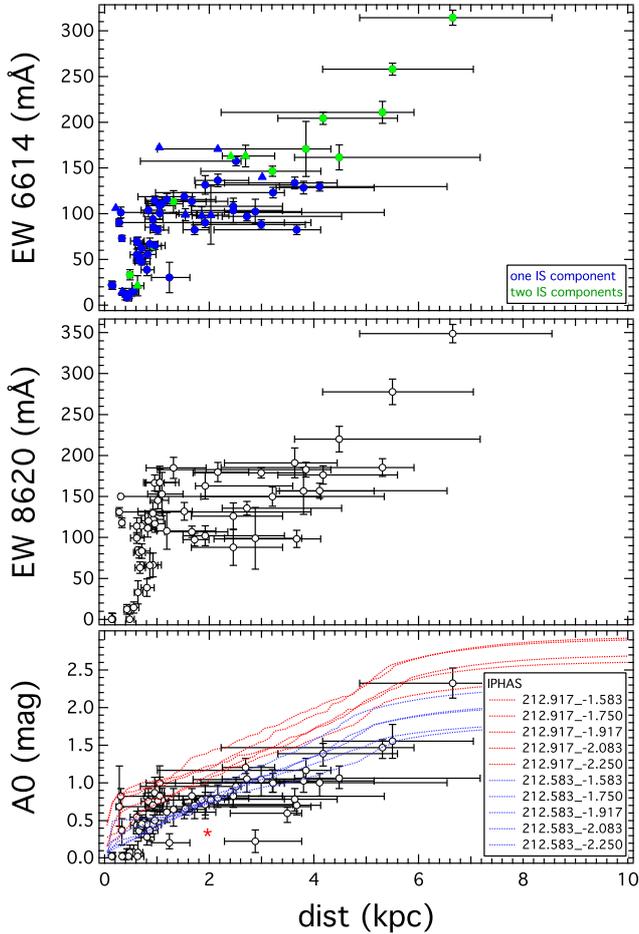}
\caption{Results for target stars in field 1 (CoRoT anticenter).  \textbf{6614 \AA\  DIB EW (top), 8620 \AA\  DIB EW (middle) and extinction $A_0$  (bottom) vs. the estimated distance}. Circles show GIRAFFE observations. Triangles show UVES observations. Colors in top panel correspond to the number of IS components used to fit the IS line or band. All nearby targets (D $\leq$ 1kpc) have only single IS component, or, for three targets, a very weak, negligible second component. Distant targets have more than one velocity component, in agreement with the crossing of at least one external arm (see the text). The outlier star 06441428-0057447 marked by (*) has stellar spectroscopic parameters in strong disagreement with stellar seismology information, which suggests they distance and extinction are inaccurate for this target.}
\label{acgraphs}
\end{figure}

For the CoRoT anti center field, the target stars are located about the Galactic Plane, and are widely distributed in distances (from 0 to 7 kpc from the Sun). This allows us to probe not only the 
local arm, but we also expect the crossing of external Galactic arms. 
As the distance of the target star increases, the LOS intersects more ISM and therefore the EW of the DIB is expected to increase, with abrupt increases corresponding to dense cloud crossings and \textit{plateaus} to interclouds/interarms.  
Figure \ref{acgraphs} shows the 6614 and 8620 bands DIB strength as well as the estimated extinction $A_0$ as a function of the target distance. We do not show the 6283 band profile due to the very limited distance range of the measurements. We remark that the DIBs and $A_0$ profiles, i.e. three quantities that are totally independently derived, are in good agreement. All of the three show a clear increasing trend, which is expected for a field of view as narrow as the one of FLAMES, 
also show the same global pattern. There is an increase between distances 0 and 1 kpc, and a second increase beyond 2.5 kpc, up to 6 kpc. These two \textit{ramps} correspond to two distinct interstellar cloud complexes, that we identify as the local and Perseus arms.  The \textit{plateau} from 1 to 2.5 kpc likely corresponds to the gap between the two Galactic arms. In this distance range there are two groups of stars with EWs that differ by about 30\%. 
They seem to correspond to two different regions within the field of view, i.e. likely the two groups do not intersect  the same parts  of the densest clouds, which is not surprising, the targets being distributed over $\sim$ 30 arcmin. Better precisions on distances and extinctions, which will be provided by Gaia, may help refine this point. 
We note a very discrepant point in the $A_0$-distance curve (marked by a red star in Fig \ref{acgraphs}, lower panel), with no corresponding anomalously small DIBs. Interestingly, this target star has sismologic parameters that are in marked disagreement with the spectrophotometric determinations (R. A. Peralta, private communication), and for this reason its distance/extinction determination may be wrong. It is encouraging that our most discrepant result points to such a contradiction.
At large distance, it is not clear whether the strong increase beyond 4 kpc corresponds to the Outer Arm. Its location is in good agreement with a crossing of the Outer Arm internal part as it appears in the schematic Galactic map of \cite{2009PASP..121..213C} (see Fig. \ref{gxmap}). Also,
the total reddening E(B-V) from the Planck map \citep{planck13} varies between 0.5 and 0.9 over the field covered by the targets ($\sim$20 x 25 arcmin wide), and the spectrophotometric extinction (or the similar DIB-based extinction) for the most distant stars is found to reach the Planck integrated value (the maximum value is even slightly above the Planck value in the direction of the corresponding target). However, we do not detect clearly a corresponding distinct and strong shift in radial velocity (see  below the discussion about the kinematics).   


\begin{figure}[h!]
\centering
\includegraphics[width=\linewidth]{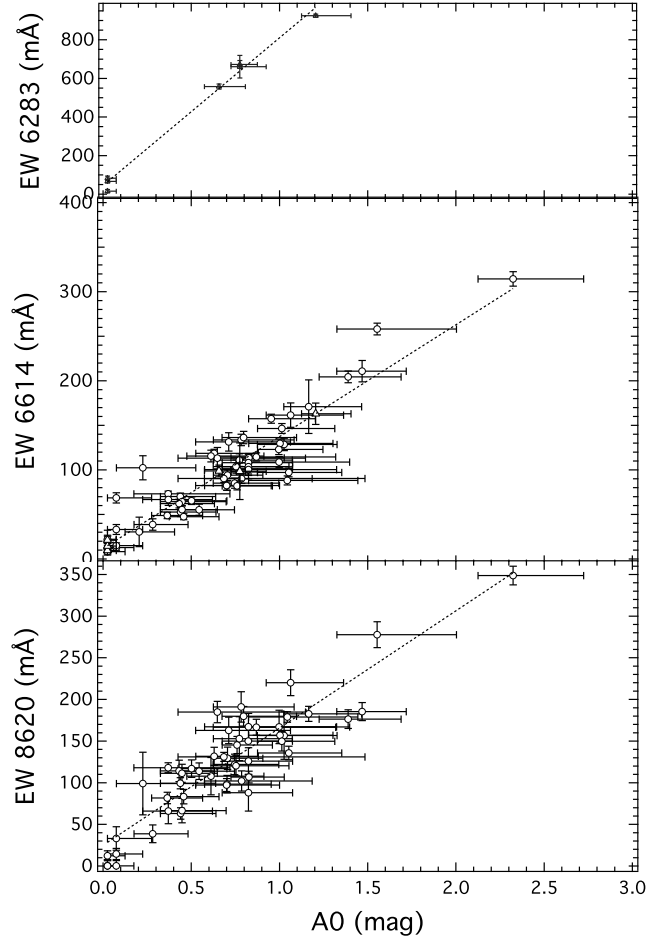}
\caption{6283, 6614, and 8620 \AA\ DIB EWs as a function of the extinction for CoRoT ANTI CENTER field targets.}%
\label{diba0corotac}
\end{figure}

Figure \ref{diba0corotac} displays, for the three DIBs, their variations with the estimated extinction $A_0$  based on all of the target stars in the field. It can be clearly seen from the figure that the three DIBs appear to be linearly correlated with the extinction.  
Our three selected DIBs are among those that are reasonably well correlated with extinction in average conditions. However, previous studies based on early-type stars have revealed a strong dispersion about the mean relationship and in particular many \textit{outliers} that correspond to the bright UV stars. Here we note that there are no equivalent \textit{outliers}, which is probably due to our cool target stars and our integrations over large distances. This corresponds to a less severely modulated character of the sight-line, or the ISM varies in a less extreme way (see the combined results in section 4.3).


\begin{figure}[ht]
\centering
       \includegraphics[width=0.42\linewidth,height=7cm]{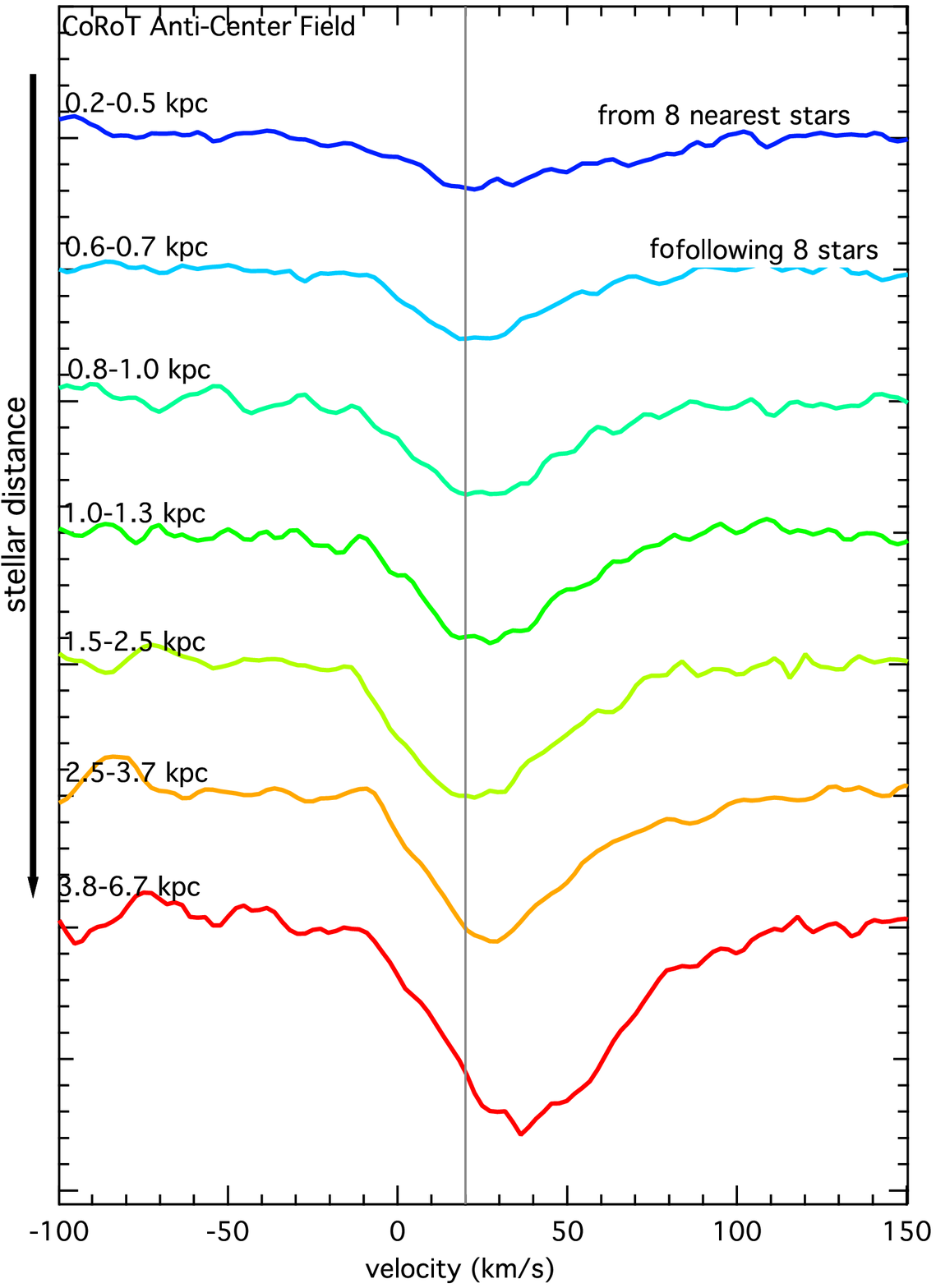}
        \includegraphics[width=0.57\linewidth]{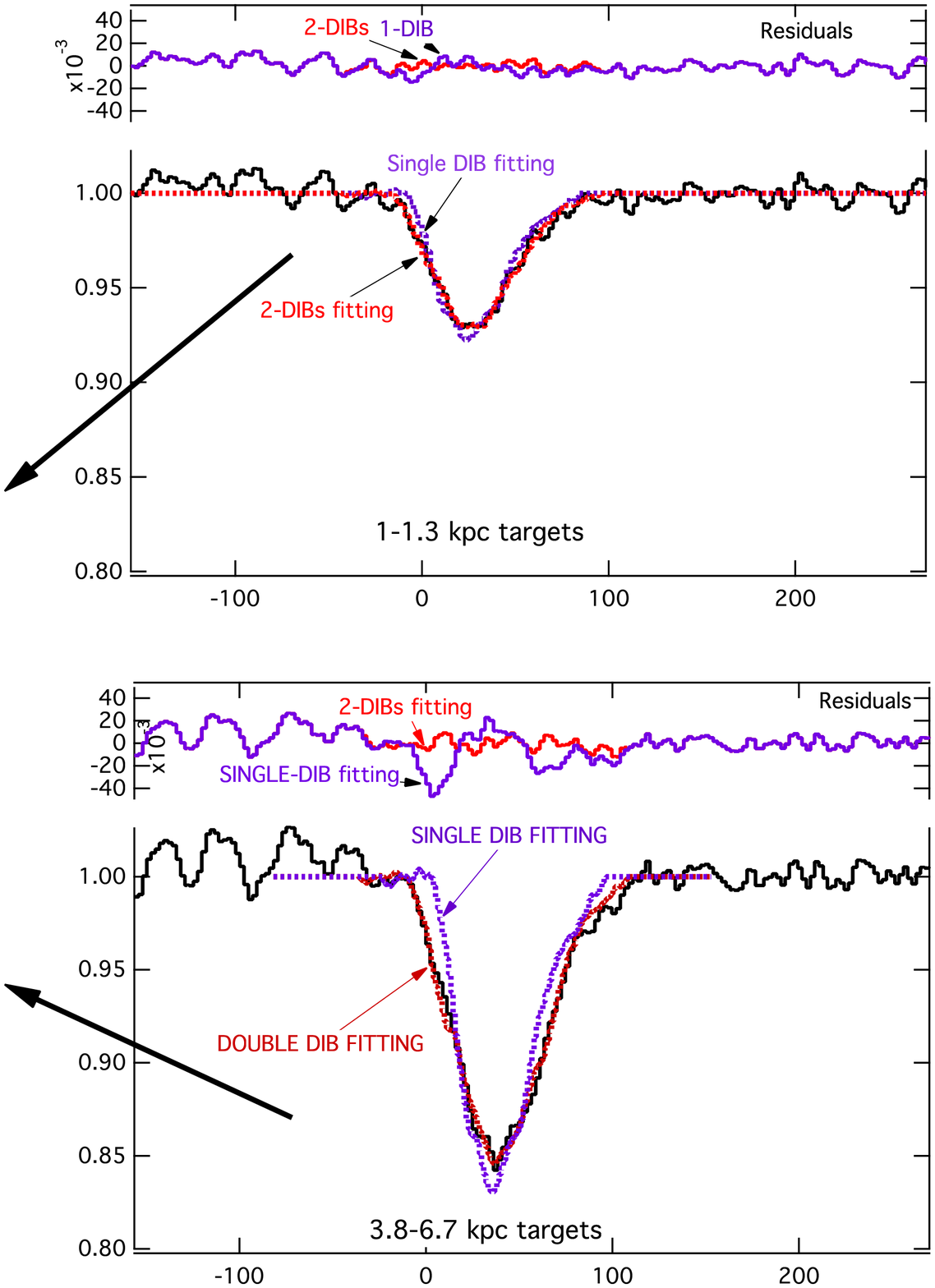}
\caption{\textbf{Evolution of the DIB profile with target distance. }Left: The 6614\AA\ DIB absorption \textbf{spectral profile} up to stars at increasing distances along the CoRoT Anti-Centre direction (l,b= 213$^{\circ}$,-2$^{\circ}$). \textbf{Shown is the average of stacked extracted, normalized absorption spectra sorted by stellar distances (an offset of 0.5  in y-axis separates two spectra}. The continuum on the blue side of the DIB is affected by the presence of strong stellar lines insufficiently corrected for. The first (top) spectrum corresponds to the first kpc, the last (bottom) spectrum to distances between 4 and 6 kpc. Right: Comparisons between single- and two-DIBs components adjustments for close and distant stars. Distant star require at least two DIBs separated by more than 20 km.s$^{-1}$ (see text).}
\label{dibevolution}%
\end{figure}


The need for a multi-component analysis in the case of the CoRoT AC field and the narrow 6614\AA\ band is illustrated in the Fig. \ref{dibevolution}. For each star we derived the full absorption attributable to the DIB in the following manner: the full profile-fitting (whose results are described below) is performed first. The fitted continuum and the adjusted stellar spectrum are used to subtract from the normalized spectrum the modeled stellar lines, leaving solely the DIB.  Within stellar line residuals, this provides the full DIB absorption independently of its assumed intrinsic shape and the number of components. After having sorted the targets by increasing distance, we averaged the absorption profiles over groups of eight stars each. The resulting profiles for each distance bin are displayed in Fig. \ref{dibevolution} as a function of the heliocentric velocity. It can be seen from the figure that the DIB depth increase with distance is accompanied with a significant velocity shift towards higher positive values, as expected from the rotation curves in this direction. The value of the maximum shift, on the order of 20 km/s, is not negligible w.r.t. the DIB width for a single cloud and calls for a multi-cloud fitting procedure. We show how this need for at least two shifted DIBs is a function of the line-of-sight extent by fitting with one then two components the mean profiles obtained from stars located between 1 and 1.3 kpc on one hand, and from stars located beyond 5 kpc on the other hand. For the most distant stars there is a strong, highly visible discrepancy between the observed profile and the adjustment with a single DIB, while the adjustment with two DIBs separated by about 30 km.s$^{-1}$ is acceptable. For the closer stars the differences between the two adjusted models are smaller and not easily detected by-eye. We have performed several statistical tests to derive the reliability of the two-DIBs model, using standard deviations derived from the continuum outside the DIB,  for both narrow or broad spectral intervals.  As a matter of fact, as we already noticed, the standard deviation varies according to the inclusion or exclusion of the spectral regions that are the most contaminated by stellar line residuals (see the blue part of the spectrum in Fig 8).  We also caution that due to those stellar lines residuals, errors on the continuum are not regularly distributed and such statistical tests are approximate, however they provide some first-order useful indications. For the distant stars the reduced chi-square increases by more than a factor of 2 when restricting to one component, i.e. this second component is statistically extremely probable, confirming the discrepancy between the observed profile and the single-DIB shape. A second test based on the Bayesian Information Criterion (BIC) is similarly showing that the existence of a second, shifted DIB is also extremely likely ($\Delta$BIC is always largely above 10). For the stars located between 1 and 1.3 kpc (middle curve in Fig \ref{dibevolution}) the reduced chi-square increase is by at least 20\%, showing that the measured profile is also very likely broadened, which is also confirmed by the BIC test. This is not so surprising as within the Local Arm the velocity dispersion may reach 20 km.s$^{-1}$. For the closest stars (top curve in Fig \ref{dibevolution}), stellar residuals in the DIB area become half of the DIB itself and a better correction of those residuals is necessary to get a firm conclusion, as confirmed by all tests.

\begin{figure}[ht]
\centering
\includegraphics[width=0.99\linewidth]{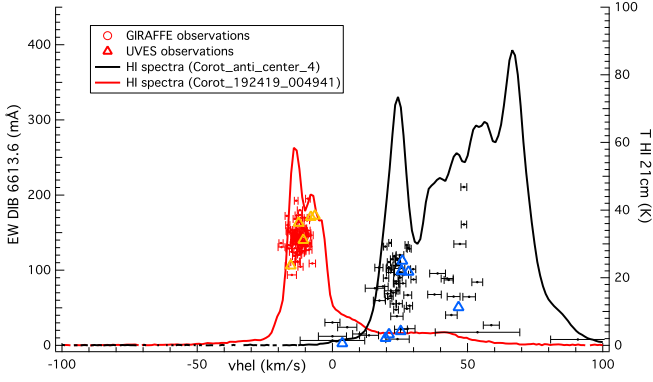}
\caption{Comparison between the fitted DIB radial velocities and EWs and the HI 21 cm emission spectra. 
Black (respectively red) markers and lines are GIRAFFE results and HI spectra from field 1 (respectively field 2). Error bars on the velocities are based on the full covariance matrix for the various parameters. In the case of the narrow 6614 \AA\  DIB (GIRAFFE observations) a significant number of spectra require two velocity components, that very likely correspond to the Local and Perseus arms. Small EWs and large error bars on velocities correspond to marginal results in low signal to noise spectra. UVES target results are displayed with triangles \textbf{(yellow and blue for fields 1 and 2 resp.)}. At variance with GIRAFFE, UVES velocities are linked to the strong sodium lines through global fitting. Their agreement with the velocity structure derived from the GIRAFFE targets shows the link between NaI and DIB velocities.
}
\label{vel6613}%
\end{figure}

We have represented in Fig. \ref{vel6613} the velocities of the detected components that come out from the automated fitting following the strategy 
described in the previous section. For all individual stars standard deviations including both the measurement uncertainties and the stellar line residuals were estimated from the best adjustments, and new adjustments were performed using these standard deviations. The errors on the free parameters were estimated using the full covariance matrix and take into account all correlations between the parameters. Resulting errors are displayed in Fig \ref{vel6613}. It can be seen that the resulting DIB velocities belong to two groups centered on:  $v_{hel} \simeq 15-32$ and $v_{hel} \simeq 40-55$ km/s (or, $v_{LSR} \simeq -2-15$ and $v_{LSR} \simeq23-38 $, respectively). Velocity results for those targets for which DIB velocities were determined through global fitting and are consequently linked to the strong sodium absorptions are marked by triangles. They are in agreement with the main groups of radial velocities, showing a global agreement between the main HI, NaI and DIB structures. The first velocity group is tightly associated with the first HI peak, that corresponds to the local arm. The second group corresponds to the second or blended second and third HI components at $\sim$ 35-45 km/s, 
that corresponds 
to Perseus.  Interestingly, none of our target requires absorption at around + 65 km/s, the heliocentric radial velocity of the reddest, strong HI emission peak (see Fig. \ref{vel6613}). It is not clear whether this highest velocity component seen in HI corresponds to the Perseus or a more distant Arm, i.e. the Outer (or Cygnus) Arm. In their synthetic Figure 3,  \cite{dame01}  are attributing to the Perseus Arm a heliocentric velocity interval $v_{hel} =37-67$ km/s ($v_{lsr} = 20-50$ km/s) in the direction of the CoRoT anti-center field, while lower velocities are predicted by \cite{vallee08}. If the higher HI velocity corresponds to the Outer Arm, then apparently none of our targets is beyond a significant column of gas/dust belonging to this Arm and all the detected IS matter is from Perseus, albeit (i) the target estimated distances are reaching at least 4.8 kpc (6.8 kpc is the most probable distance), (ii) there is a strong and coherent  increase of DIBs and extinction with distance found from the 6 most distant targets, and, (iii) as discussed above the Planck integrated reddening is on the same order than the reddening towards our most distant targets. In this case the  fastest HI arises beyond 5-6 kpc, and is too poor in dust to produce a significant additional reddening. Conversely, if the faster gas belongs to Perseus, a potential explanation is that the ensemble of distant targets may miss those clouds. HI maps have a lower resolution compared to Planck, the Perseus Arm is highly fragmented and the distant targets are distributed over $\sim$ 15 arcmin. If there is a strong inhomogeneity within the field the  path to the distant  targets may not cross the higher velocity matter. More data are needed and more accurate distances should help answering this question.


\begin{figure}[ht]
\centering
	\includegraphics[width=0.5\linewidth]{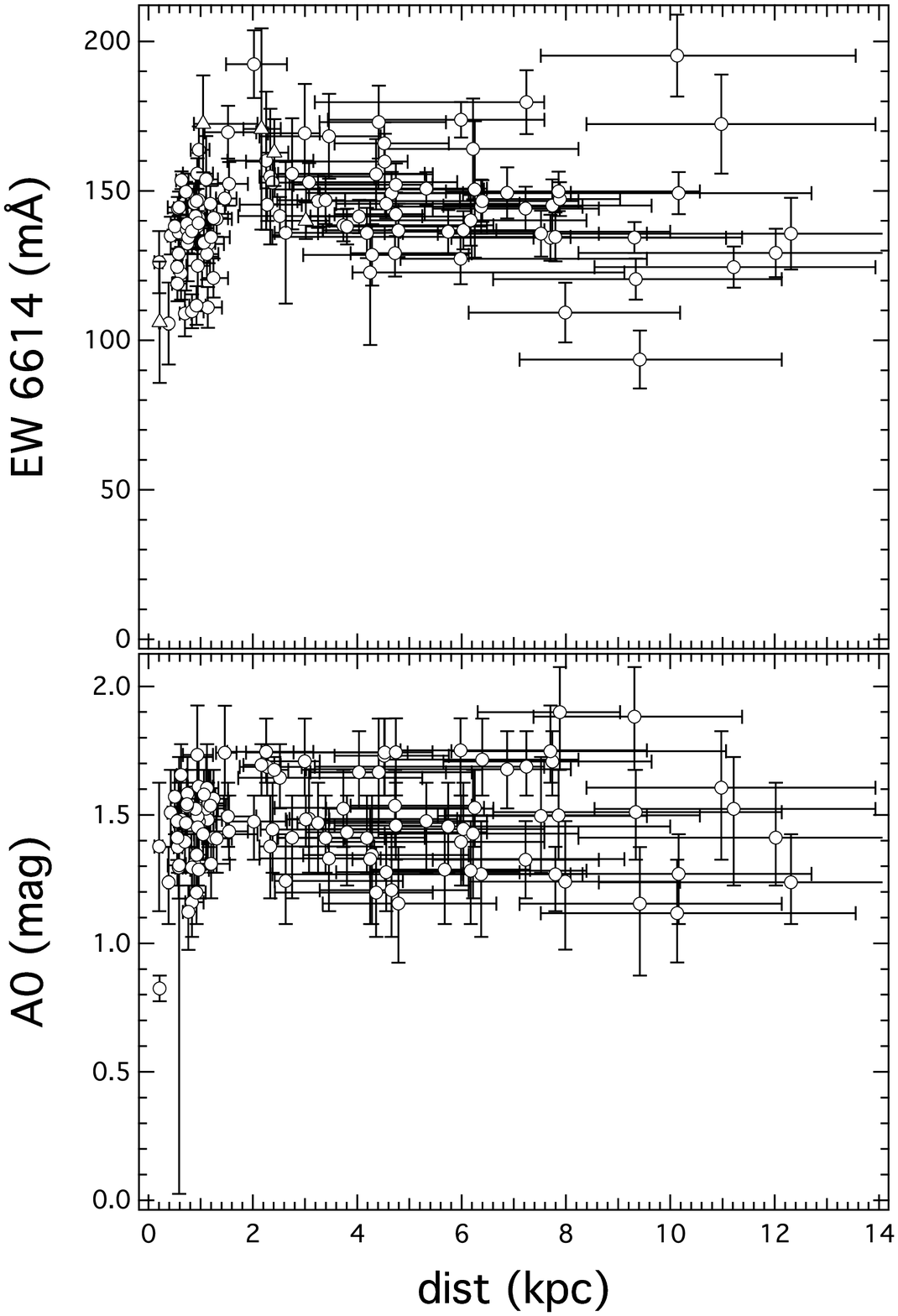}
	\includegraphics[width=0.5\linewidth]{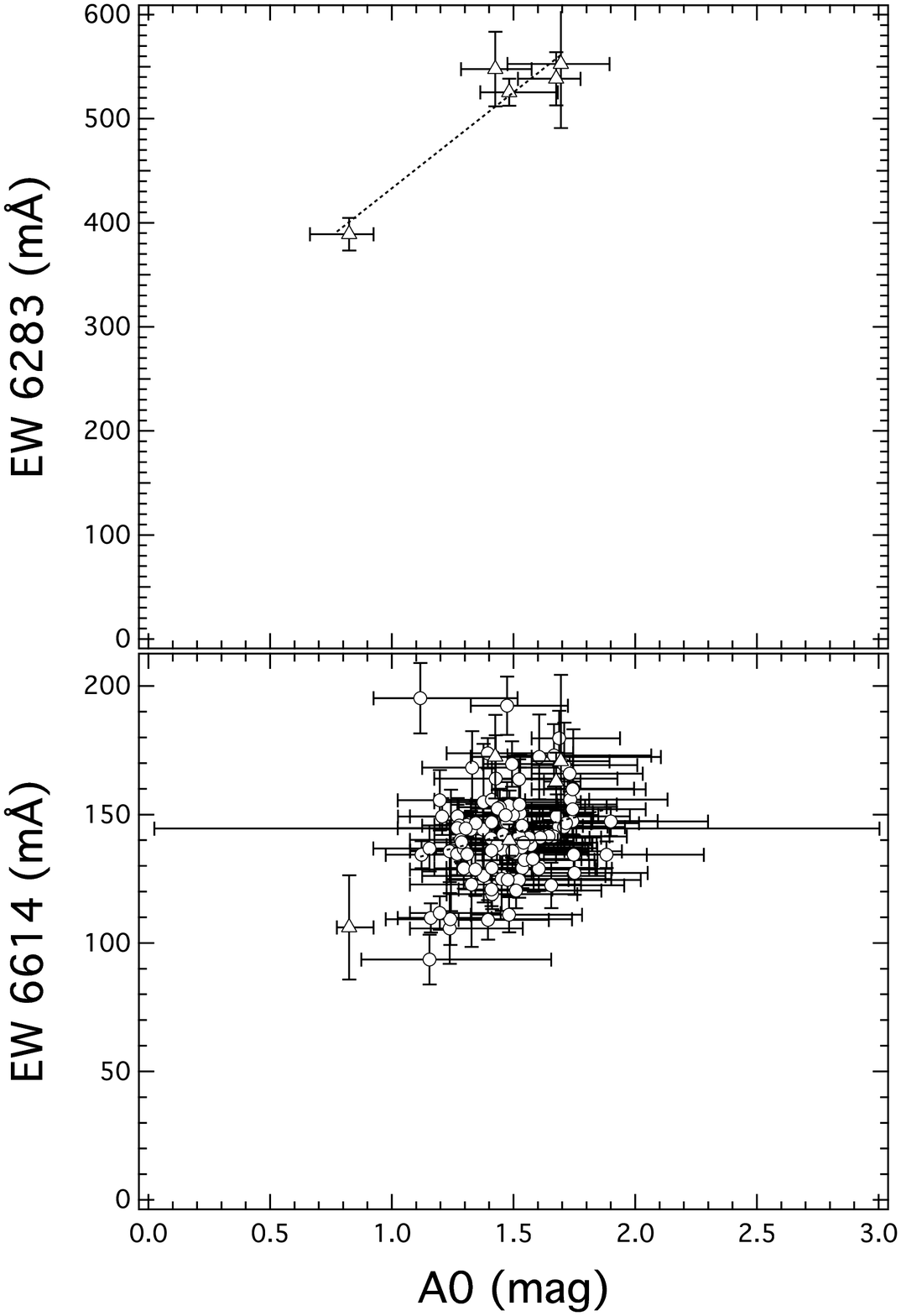}
\caption{CoRoT CENTER FIELD field. Left:  DIB/$A_0$ vs. distance profile. Right: DIB vs $A_0$: 6283 and 6614}
\label{COROTC}%
\end{figure}

For the CoRoT center field, target stars also widely distributed in distance, however, 
its  higher latitude ($b=-7^\circ$) has a strong impact on the results. For this field,  the 8620 \AA\ DIB is not extracted due to significant sky emission line residuals. 
Figure \ref{COROTC} displays the DIBs and the extinctions as a function of the target distance. Although distributed over large distances
, we do not detect any  EW increase (\textit{ramp}) in addition to the one associated with the local arm. 
Instead, the DIB strength appears to form a \textit{plateau}. 
This shows that the LOS do not intersect inner arms 
because the distant target stars are significantly below the Plane. The measured profile implies that most of the absorbing matter is closer than 1.5 kpc. 
The relationships between the DIB strength and the extinction is shown in Fig. \ref{COROTC}. Due to the quite small DIB and extinction interval covered by the targets in this field, all of the data points are clustered. Still, an increasing trend is clearly observed. For this field, 
the kinematics is also rather simple (Fig. \ref{vel6613}). 
There was no need for more than one IS component, all velocities fall close to each other in agreement with the peak of the HI emission spectrum at  $v_{hel} \simeq -14$ km/s. 

\subsection{Fields 3 to 7}

\begin{figure}[t]
\begin{center}
\includegraphics[width=0.49\linewidth]{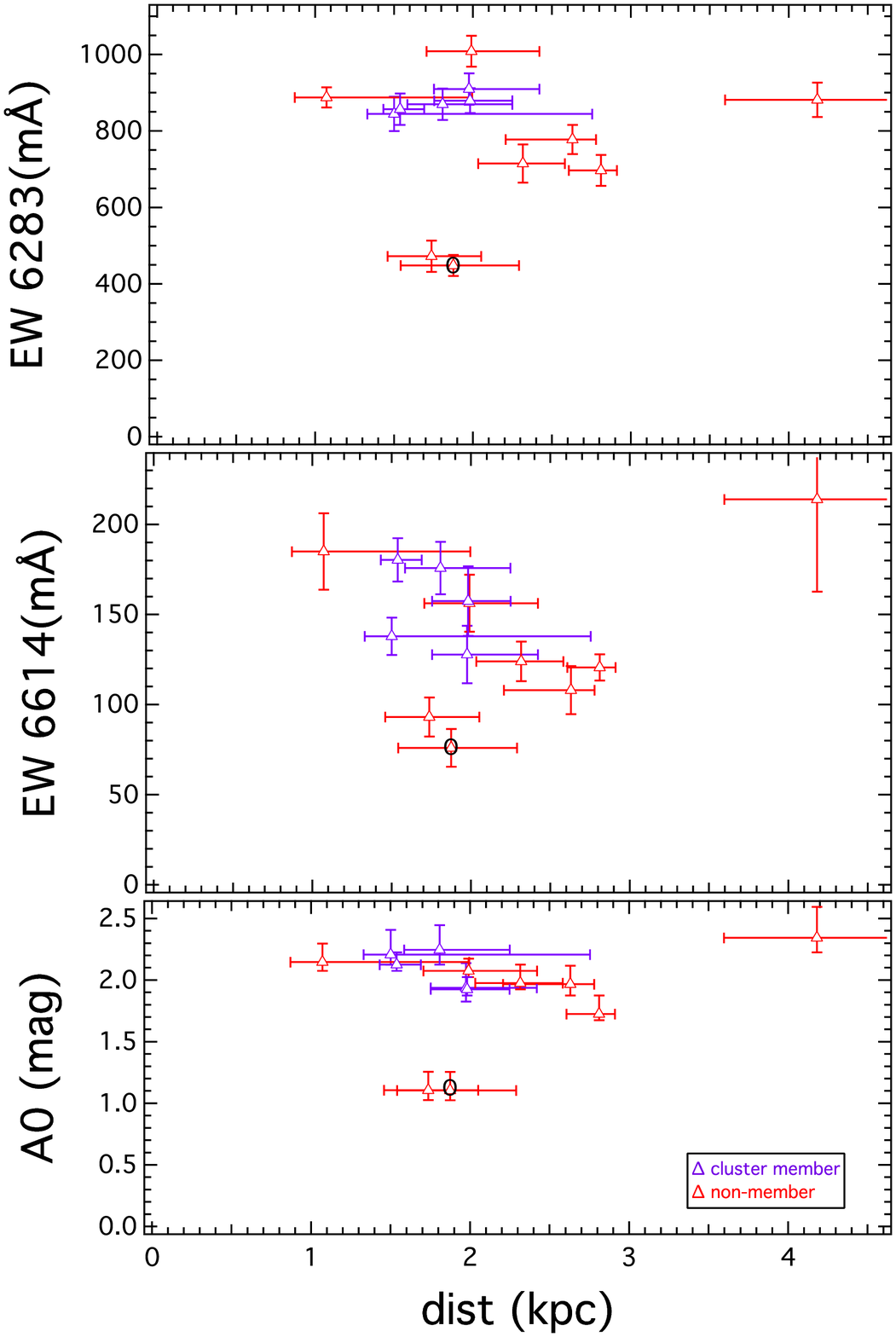}
\includegraphics[width=0.49\linewidth]{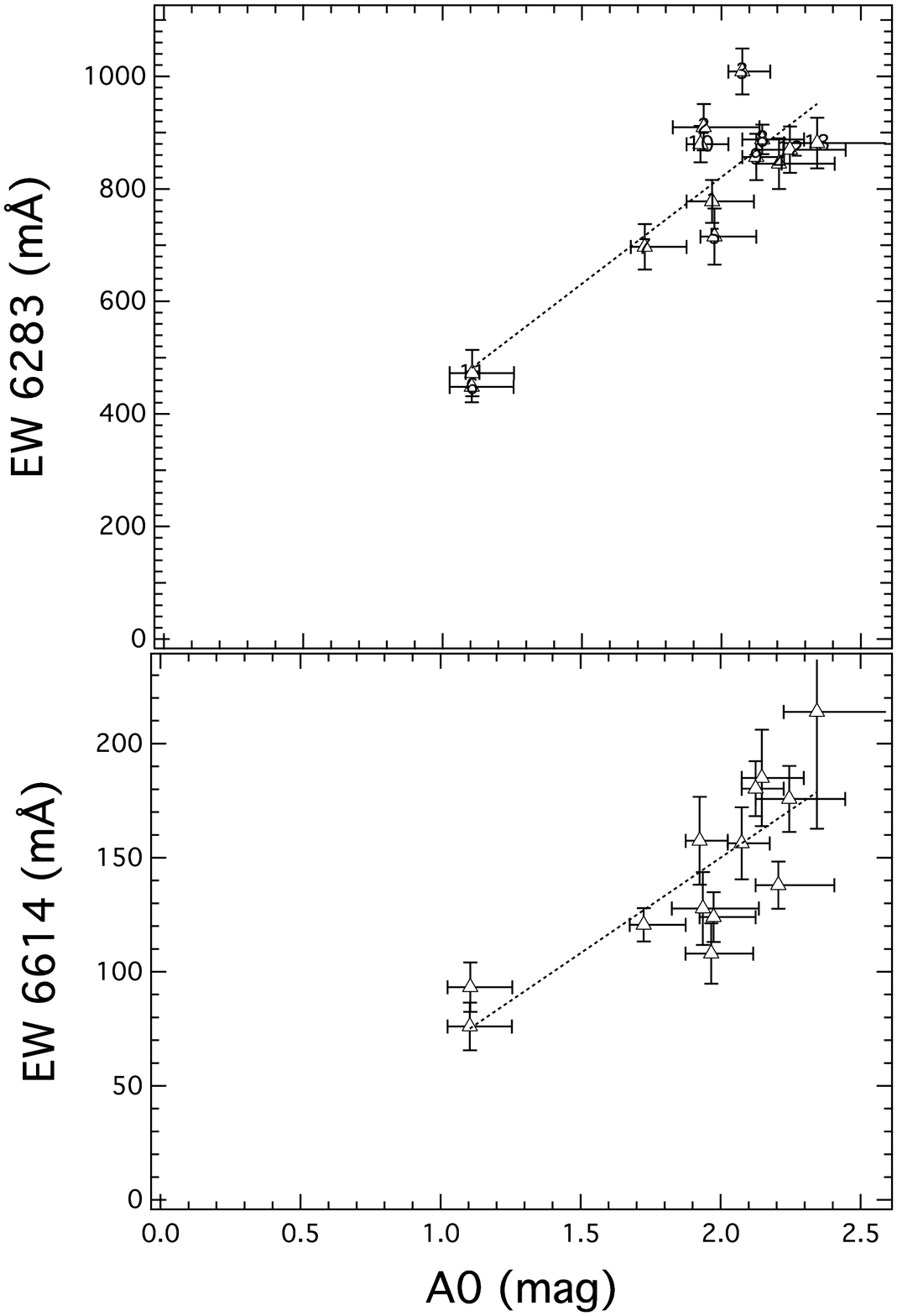}
\caption{NGC 4815 field: (left) DIB and $A_0$ distance profiles. Stars identified by \cite{Friel} as cluster members correspond to blue markers, non-members to red. Right:  6283 (top) and 6614 (lower panel) \AA\ DIB EW vs the estimated extinction A0. The black "0" sign indicates the unique star with a single DIB velocity, for all other targets adjustment to data requires two velocity components.}
\label{dibdistcluster}
\end{center}
\end{figure}

\begin{figure}[htbp]
\begin{center}
\includegraphics[width=\linewidth]{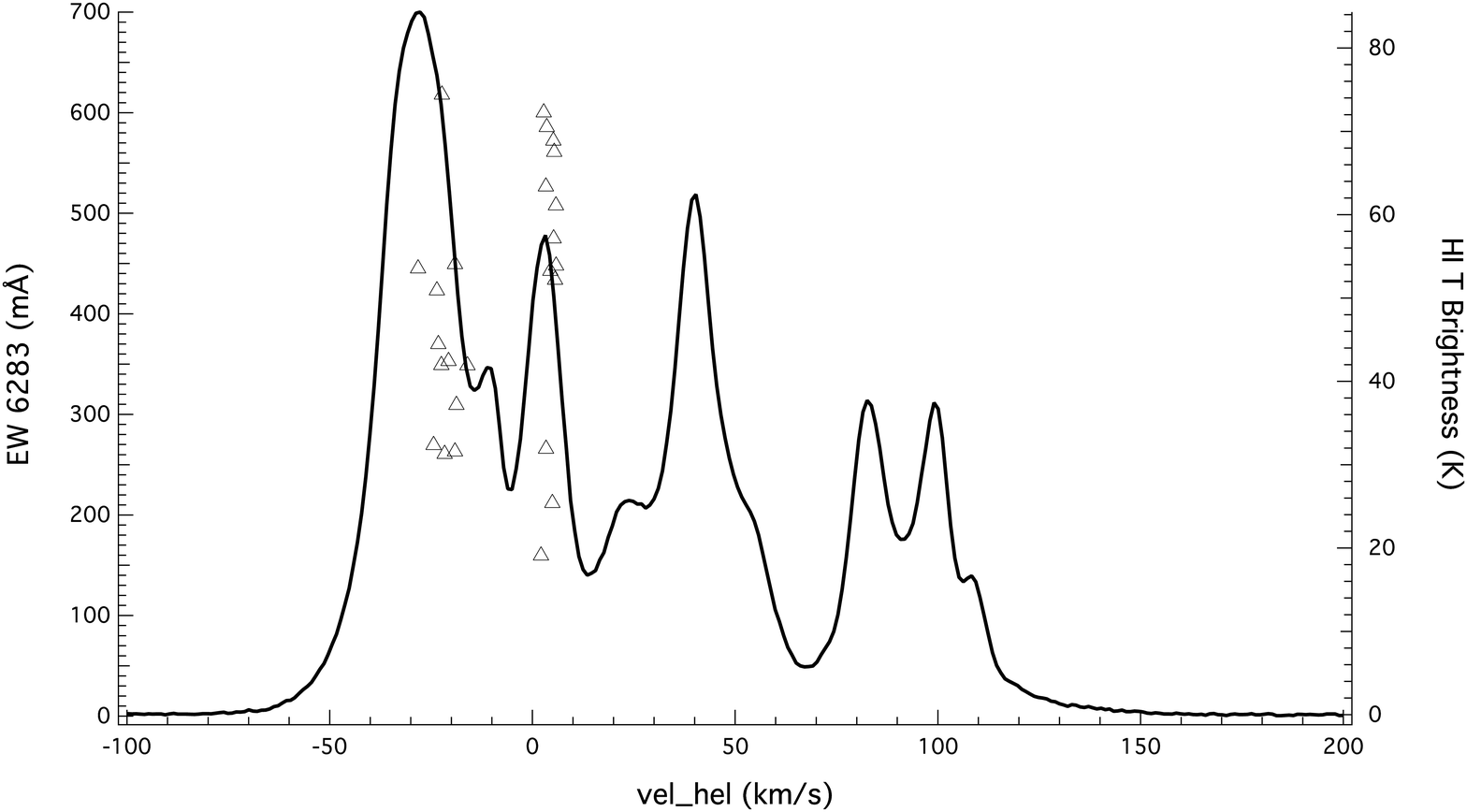}\\
\includegraphics[width=\linewidth]{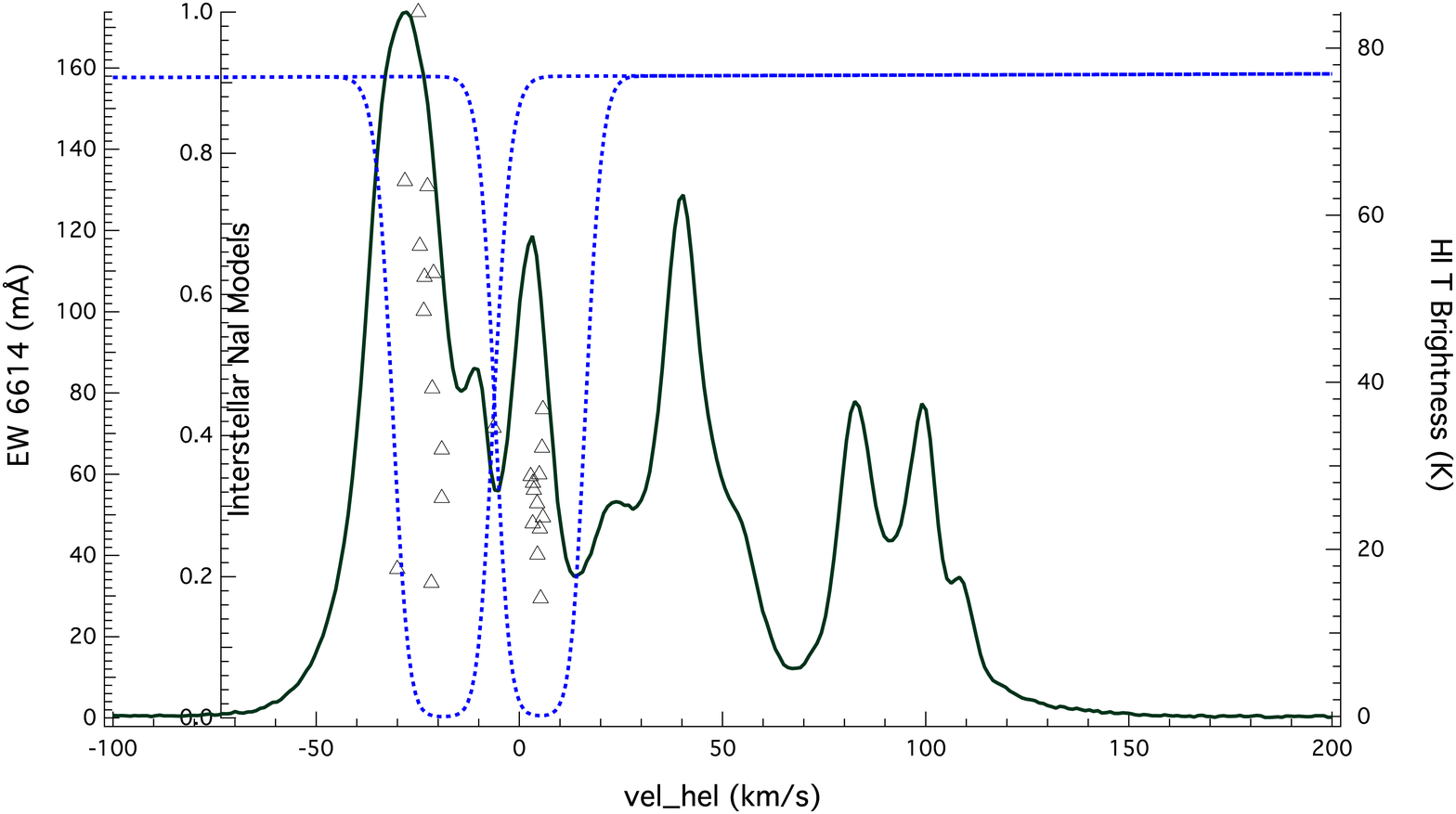}
\caption{NGC 4815 field: kinematics. Top: the 6283 \AA\ DIB. Bottom: the 6614 \AA\  DIB. The black line represents the HI 21 cm emission spectrum (LAB Survey). The dashed blue lines are an example of the fitted IS NaI lines (here from star cname 12581939-6453533) and triangles are the velocities of the DIB components derived from the global fit for all targets.}
\label{velcompcluster}
\end{center}
\end{figure}

For the field3/NGC\,4815 direction  \citep[see][]{Friel,Magrini}, 
we analysed 14 red clump stars that were observed with UVES. Only six are open cluster members. 
We performed the simultaneous fitting of the NaI lines and the 6283  \AA\ DIB (respectively the 6614 \AA\ DIB). 
The NaI absorption is characterized by two velocity components, at $v_{hel} \simeq$ -25 and -5 km/s (see Fig. \ref{velcompcluster}), that are well separated and was useful to test our multi-component technique. 
The results are displayed in Fig. \ref{dibdistcluster}. 
The extinctions and DIB strengths are on the same order for most stars, 
showing that at their distances on the order of 2 kpc,  they are located beyond the main, nearby absorber, in agreement with the 3D ISM map \citep{2014A&A...561A..91L}. 
The two stars with lower extinction are not cluster members and
must be foreground stars. Their most probable distances are 1.7 and 1.9 kpc, which shows that not all the absorption is local 
and pinpoints another absorber 
between 1.9 kpc and 2.5 kpc (cluster distance). 
From the results of the most distant target, 
there is no significant additional IS absorption between 2 and 4 kpc. 
The comparison between the DIBs and the estimated extinctions shows they are well correlated (Fig. \ref{dibdistcluster} right). 
%
Radial velocities of the NaI lines and DIBs correspond to two strong peaks in the HI spectrum. Those cloud complexes also appear in the CO survey of  \citet{1987ApJ...322..706D} and probably correspond to the Coalsack complex and another dense cloud. In none of the spectra do we detect velocities 
above +10 km/s, which implies that 
those HI structures  at higher velocity are located beyond 4 kpc.



\begin{figure}[htbp]
\begin{center}
\includegraphics[width=0.5\linewidth]{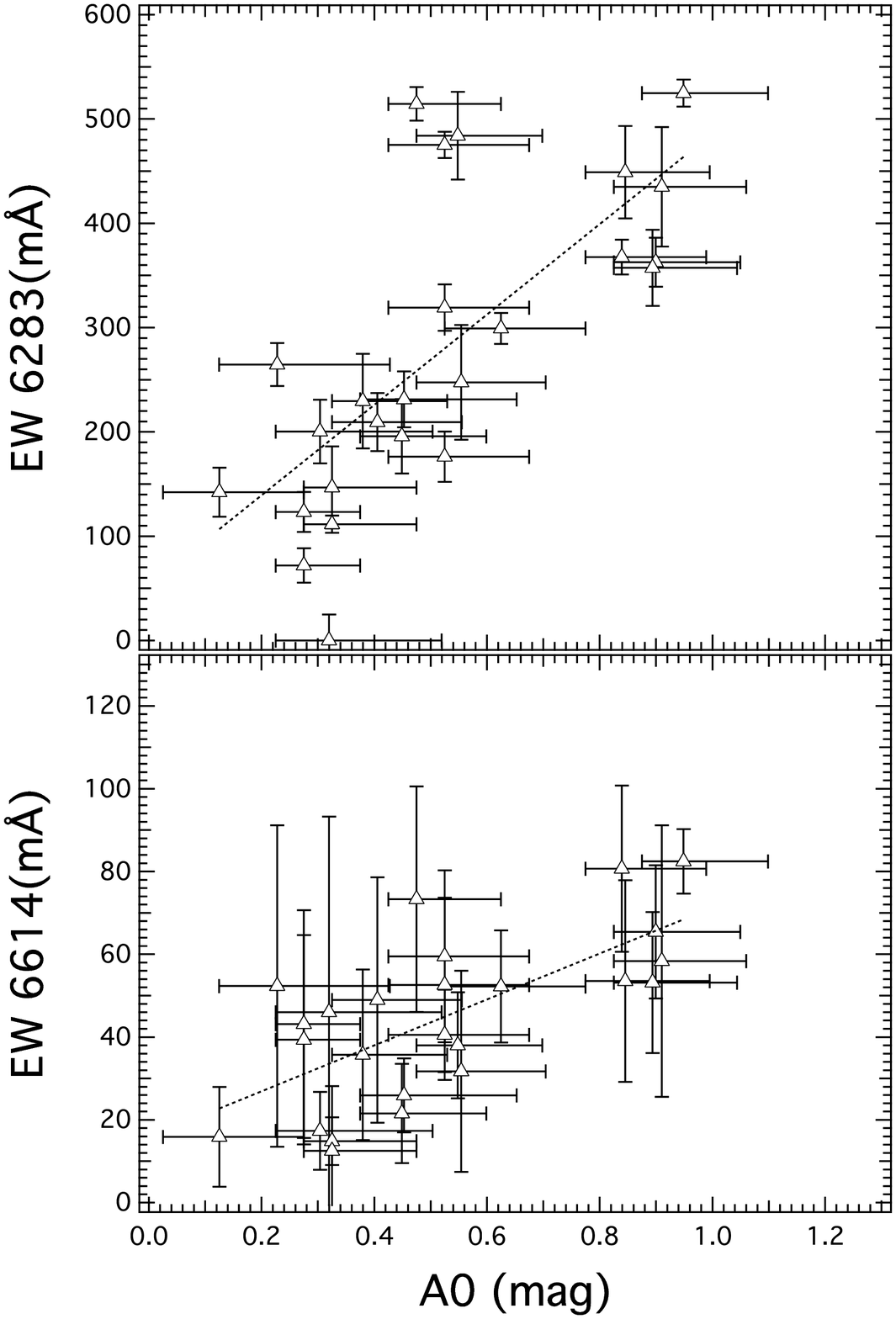}
\includegraphics[width=0.5\linewidth]{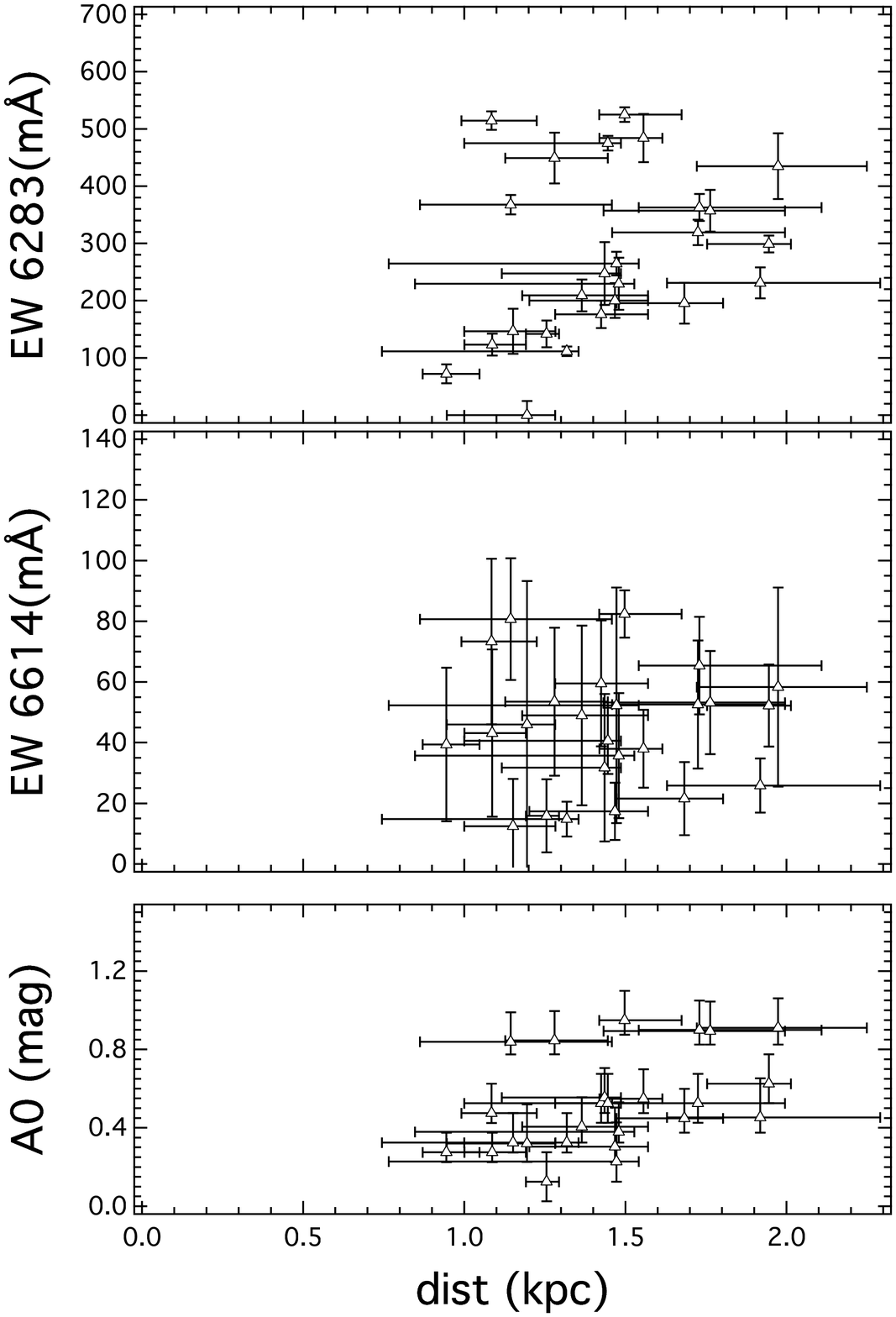}
\caption{The  $\gamma$ Vel field. Left: DIB EW and estimated extinction as a function of target distance. Two groups of stars are present that probe different regions of the foreground cloud. Right: The relation between EWs and  extinction. There is a significant scatter, the largest from the 7 fields. Several \textit{outliers} have stronger DIBs compared to the averaged relation. These departures from linearity are very likely linked to the influence of the Wolf-rayet star environment.}
\label{dibdistgam}
\end{center}
\end{figure}

\begin{figure}[t]
\begin{center}
\includegraphics[width=\linewidth]{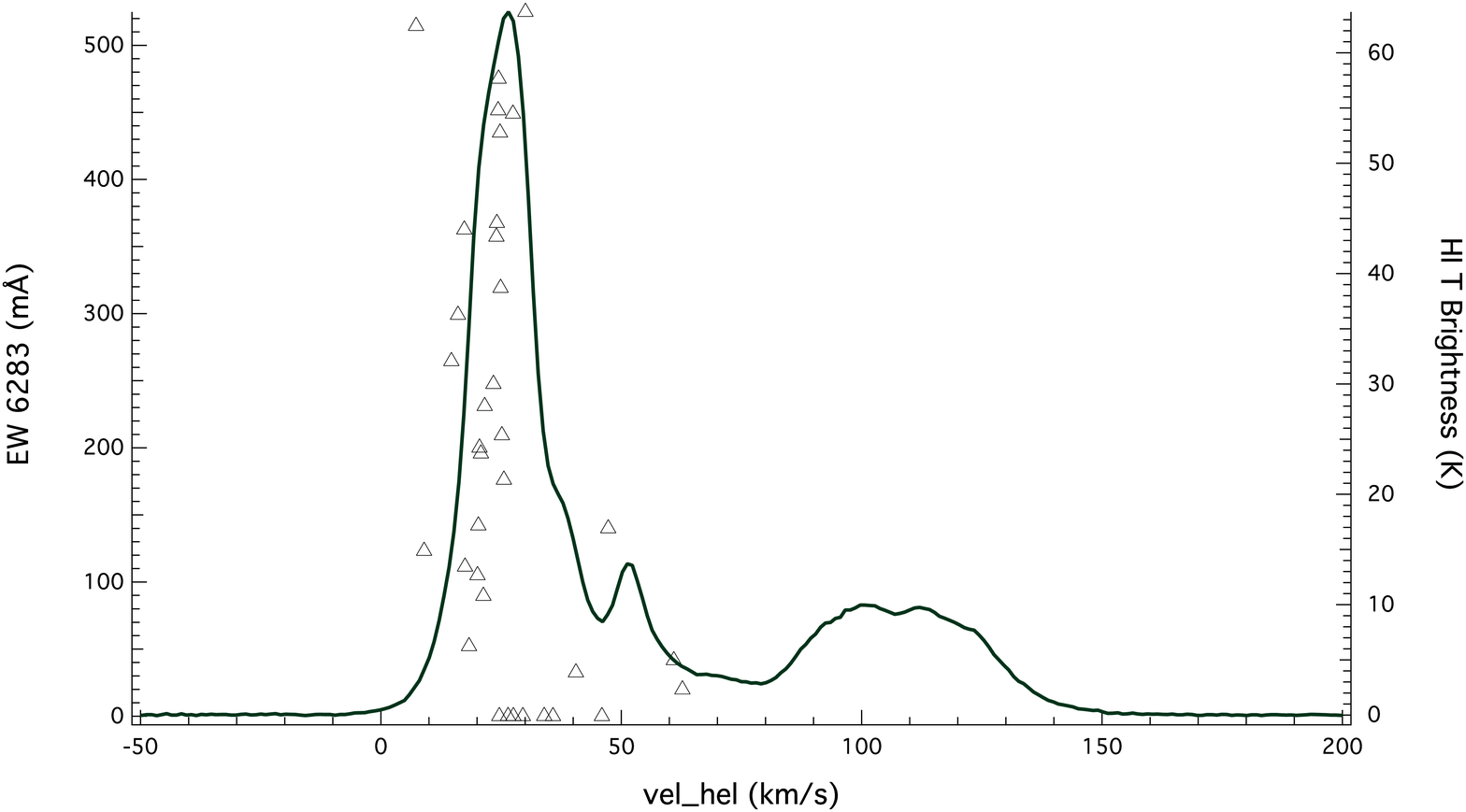}\\
\includegraphics[width=\linewidth]{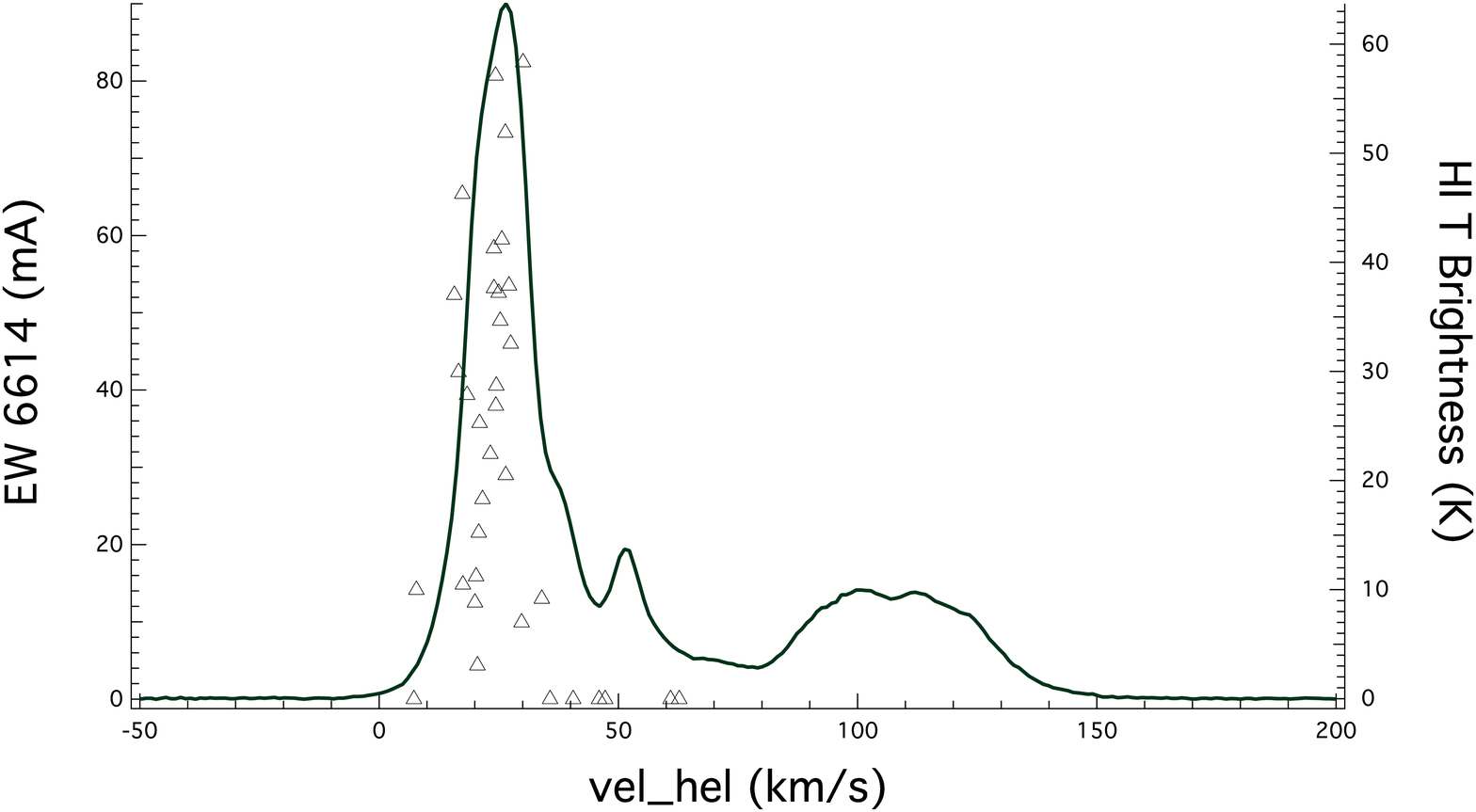}
\caption{$\gamma$ Vel field: kinematics. Top: the 6283 \AA\ DIB. Bottom: the 6614 \AA\ DIB. Line and markers are the same as the one in Fig. \ref{velcompcluster}.}
\label{velcompgam2vel}
\end{center}
\end{figure}

For the $\gamma$ Vel direction (field 4)  \citep[see][]{jef09,jef14,spina}, 
the most significant difference from the other directions is the distribution of targets over a much wider area ($1^{\circ} \times 1^{\circ}$). 
As we will see (in Figure \ref{dibdistgam}) this has a strong impact on the star to star variability, especially for this region that is well known for having a complex interstellar density and ionization structure, partly under the strong influence from the Wolf-Rayet  (WR) star.  
For the 6614 \AA\ DIB, unfortunately, the profile is significantly scattered due to the relatively strong influence of the stellar residuals and the resulting large relative errors. 
The HI spectrum presents a strong peak at $v_{hel}$ 35 km/s, a velocity that is in good agreement with the NaI and DIB absorptions (see Fig. \ref{velcompgam2vel}). 
The second component in the HI spectrum at $v_{hel}$ 50 km/s is found to be very weak or null.  Whereas, the HI component at $\simeq$ 100 km/s is not detected in any of the spectra and corresponds to more distant clouds.

\begin{figure}[htbp]
\begin{center}
\includegraphics[width=0.5\linewidth]{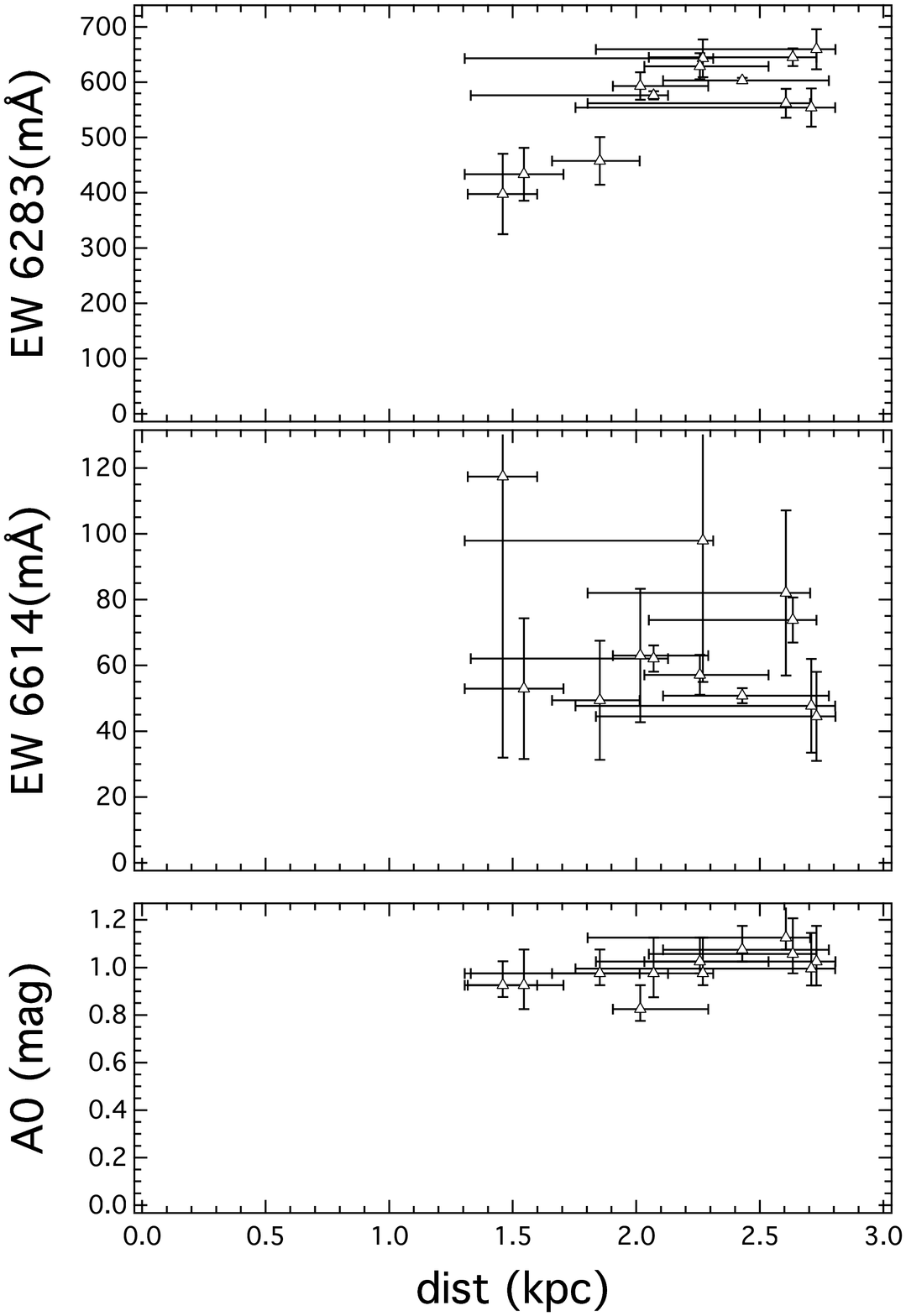}
\includegraphics[width=0.5\linewidth]{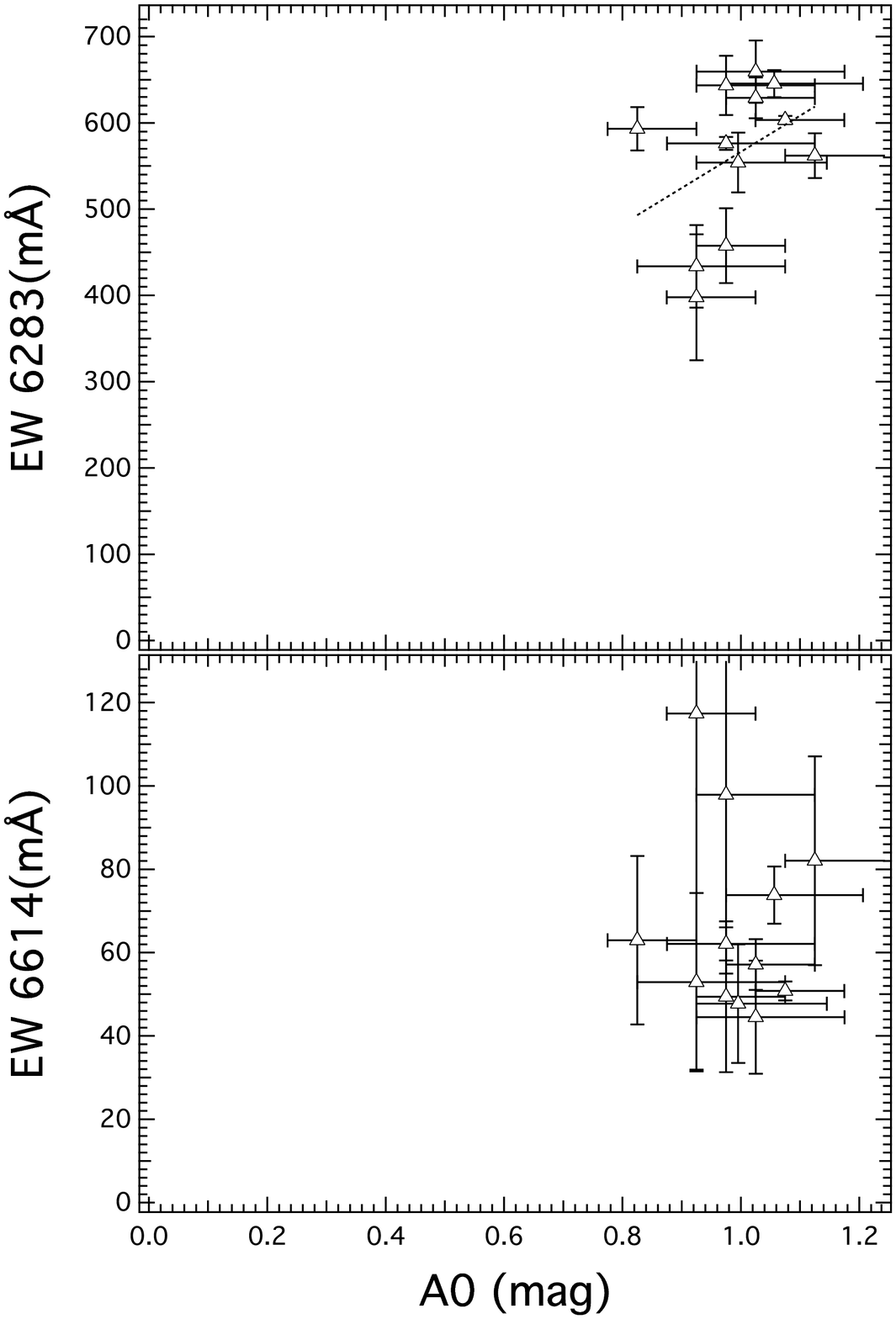}
\caption{Baade Window direction: DIB vs. distance and $A_0$ vs. distance}
\label{dibdistbulge}
\end{center}
\end{figure}

\begin{figure}[htbp]
\begin{center}
\includegraphics[width=\linewidth,height=4.cm]{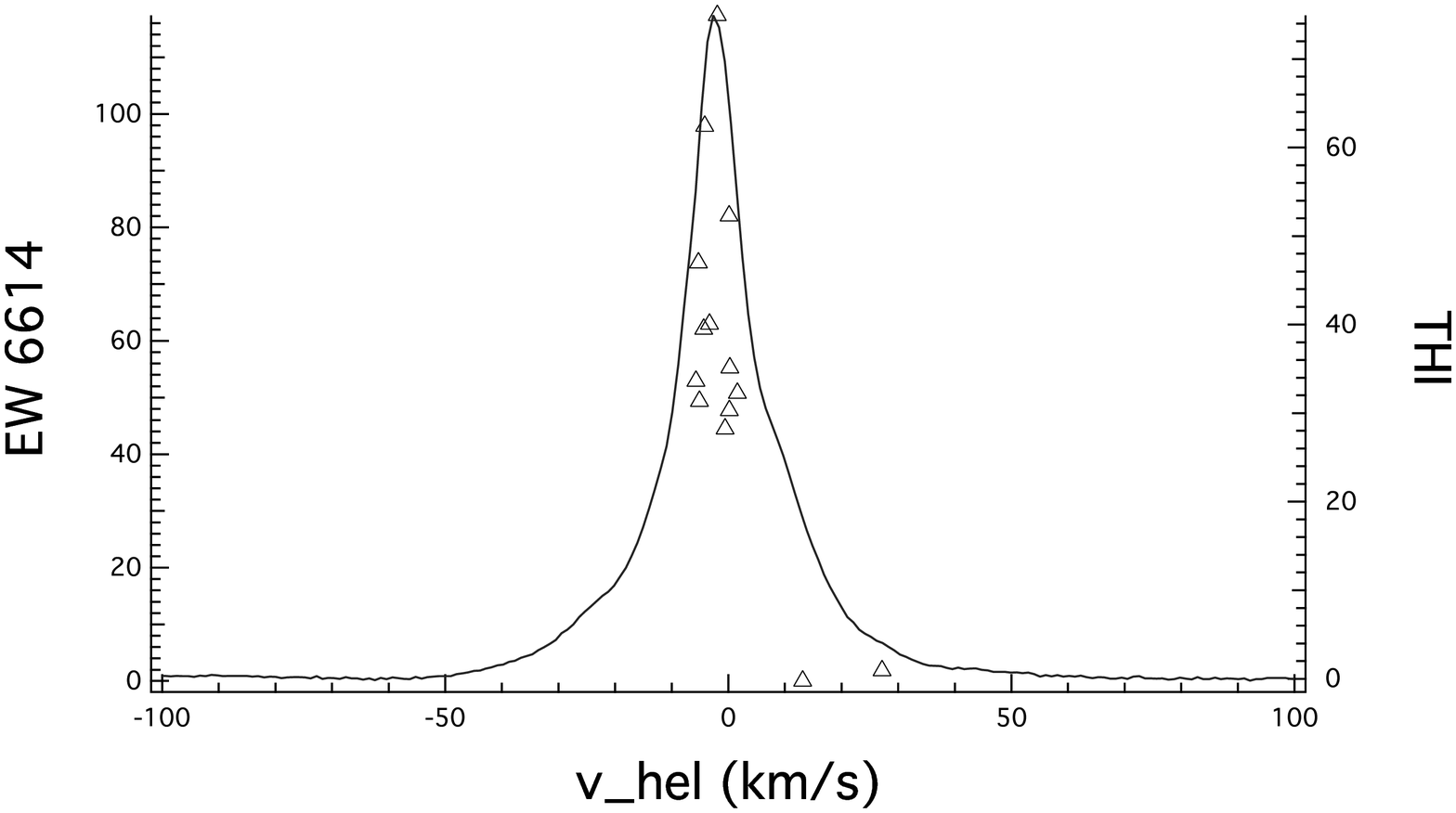}\\
\includegraphics[width=\linewidth,height=4.cm]{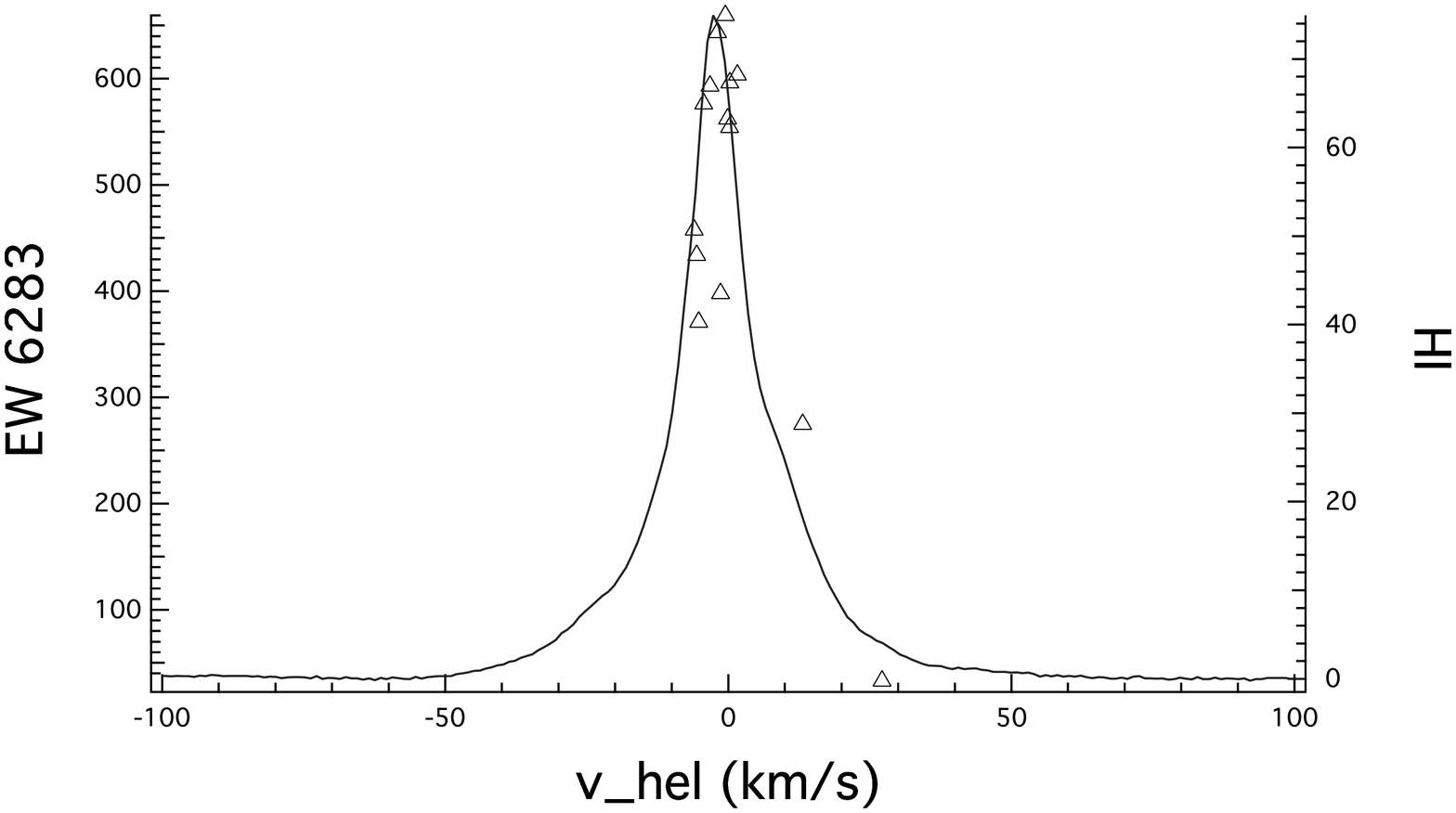}
\caption{Baade Window direction: kinematics. All DIB velocities are found to be consistent with the local HI, around 5 km/s. The second velocity component allowed by the fitting procedure  is found to be unnecessary or negligible. Note that the discrepant data point at +15 km/s for the 6283 \AA\ DIB (lower panel) is due to the effect of a strongly discrepant stellar line and disappears when the fitting is repeated after masking of the corresponding region (the total EW is found to be unchanged).}
\label{velcompbulge}
\end{center}
\end{figure}

The last three fields point to the Galactic bulge. This first field corresponds to the commonly used, low extinction direction at $(l,b)\simeq(1^\circ,-4^\circ)$, the Baade's window (BW). 
Figure \ref{dibdistbulge} (left) shows the extinction, and the 6283 and 6614 EWs 
along this LOS. 
As for the previous field, the 6614 \AA\ profile is consistent with the two others, however much less precisely defined because the absorption is weaker and the stellar line residual have a stronger impact. 
Figure \ref{velcompbulge} shows the HI emission spectrum in this Bulge direction, a spectrum characterized by a dominant emission peak at -5 km/s  (heliocentric frame). Here again, the comparison with the DIB velocity that comes out from the automated fitting shows a good agreement with HI, with a dispersion of a few km/s around the central velocity. 

\begin{figure}[htbp]
\begin{center}
\includegraphics[width=\linewidth]{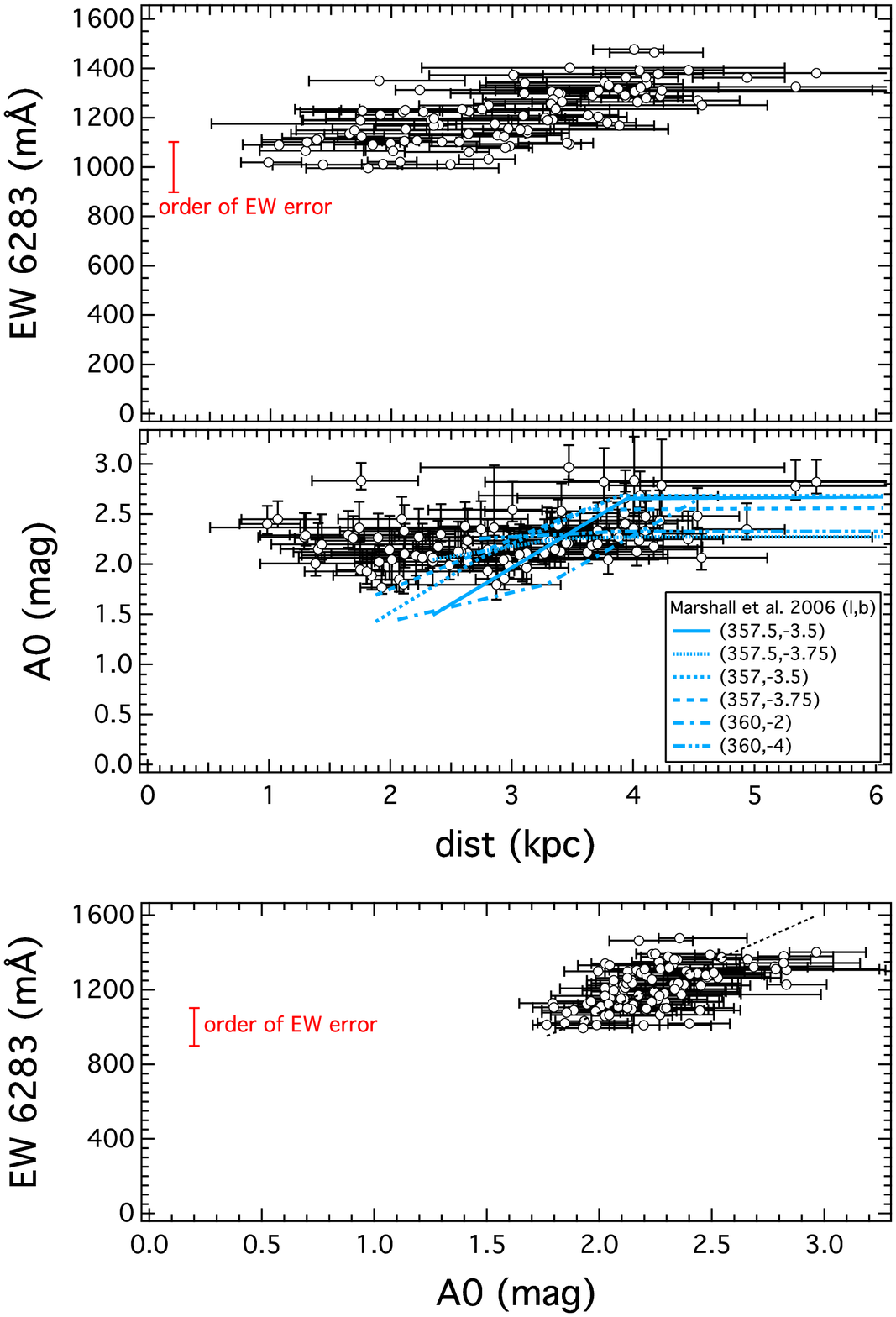}
\caption{6283 DIB EW and estimated extinction as a function of distance (top) and DIB extinction relationship  (lower panel) for the OGLE BUL\_SC24 (INNERDISK O) field. We compare our estimated extinction with the profiles from \cite{marshall06} (see text) for \textbf{the closest} directions.}
\label{dibdistinnerdisko}
\end{center}
\end{figure}

\begin{figure}[htbp]
\begin{center}
\includegraphics[width=\linewidth]{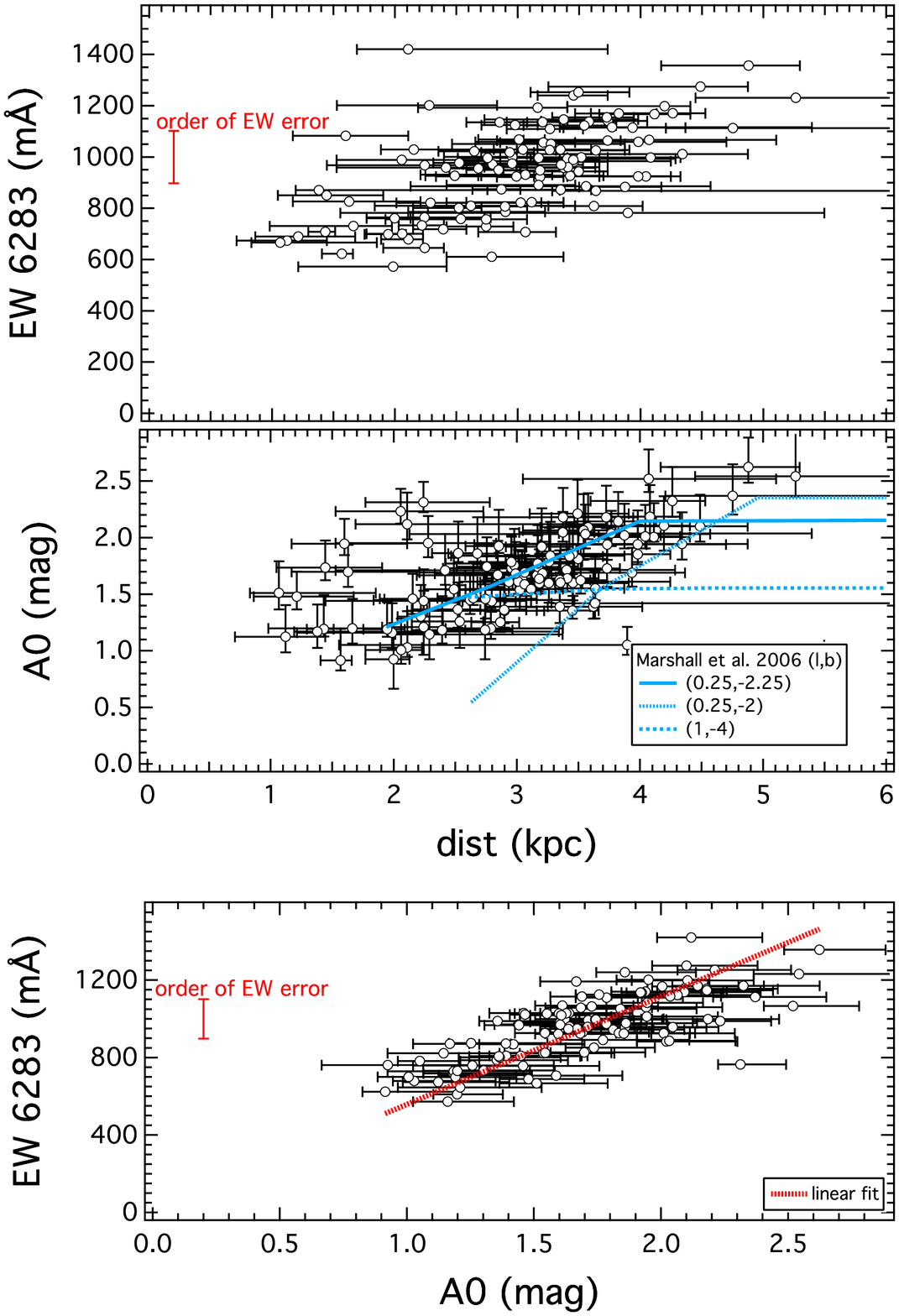}
\caption{INNERDISK W field: DIB vs. distance, $A_0$ vs. distance, and DIB vs $A_0$. \textbf{The EW vs $A_0$ linear relationship (lower panel) has a slope of 567 $\pm$ 7 mA per magnitude (forcing the intercept to be 0). The Pearson correlation coefficient is 0.76.} }
\label{dibdistinnerdiskw}
\end{center}
\end{figure}

For the next two fields  
(INNERDISK O and W resp.), 
IS absorptions are expected to be confined within a narrow interstellar radial velocity range, which is confirmed by the HI emission spectra, and the DIBs could be analyzed by means of the single component method. We have checked for several stars that  the allowance for more than one component results in  EW values that are fully compatible with those from the single component method, within our estimated uncertainties.
Figure \ref{dibdistinnerdisko} and \ref{dibdistinnerdiskw} show the radial profiles of the DIB strength and the estimated extinction. They both show a gradual  increase. 
The profiles are in agreement with the profiles derived by \cite{marshall06} from 2MASS and the Besançon model in adjacent directions. We used A$_{Ks}$/A$_{0}$=0.11 for the conversion. The DIB-extinction correlation is shown in lower panel, and is compatible with a linear relationship within the measurements and the model uncertainties. The Pearson coefficients are found to be 0.55 and 0.76 for INNERDISK O and W resp.


\subsection{All fields: Correlation with the Extinction}

\begin{figure}[h!]
\centering
	\includegraphics[width=\linewidth]{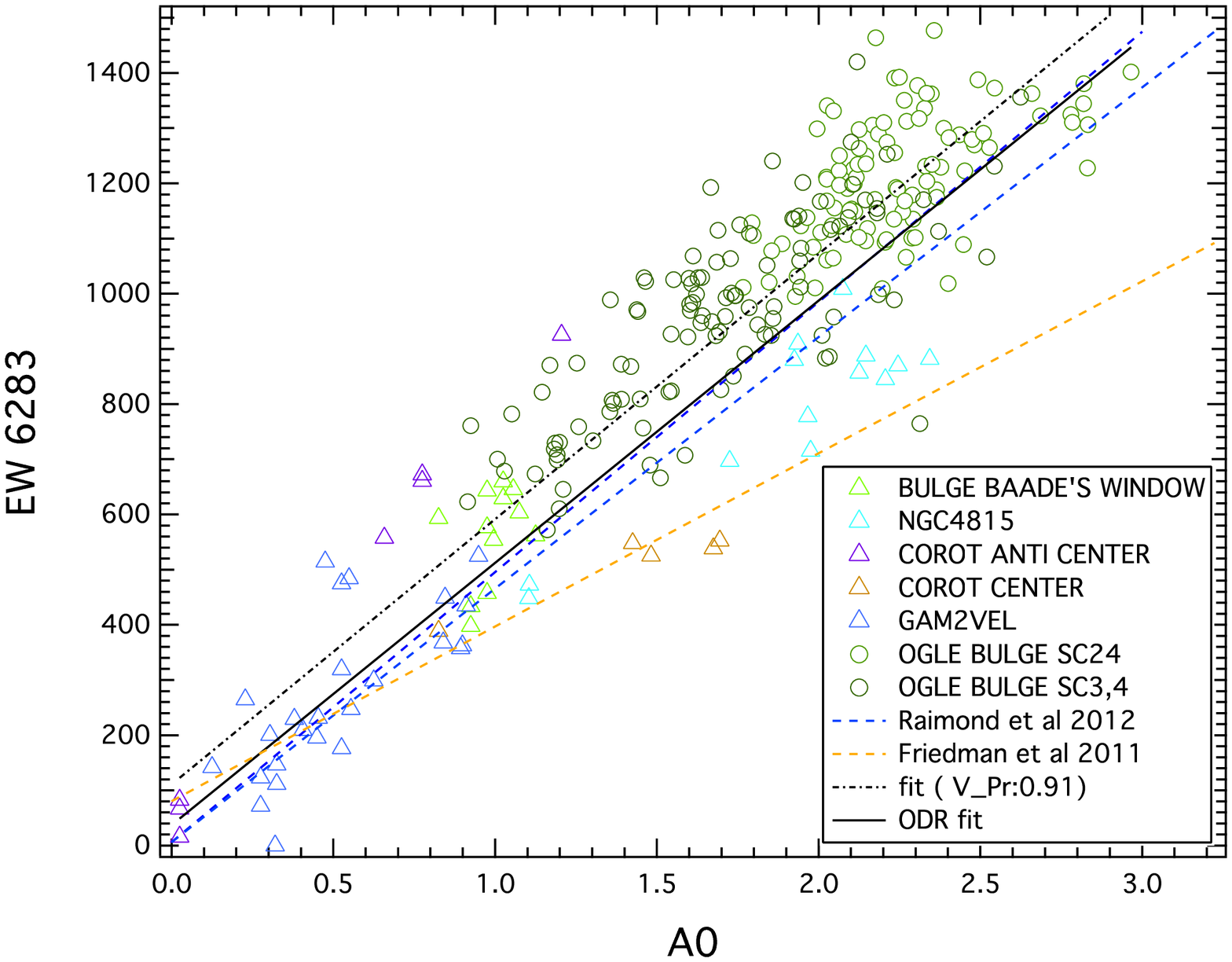}\\
\includegraphics[width=\linewidth]{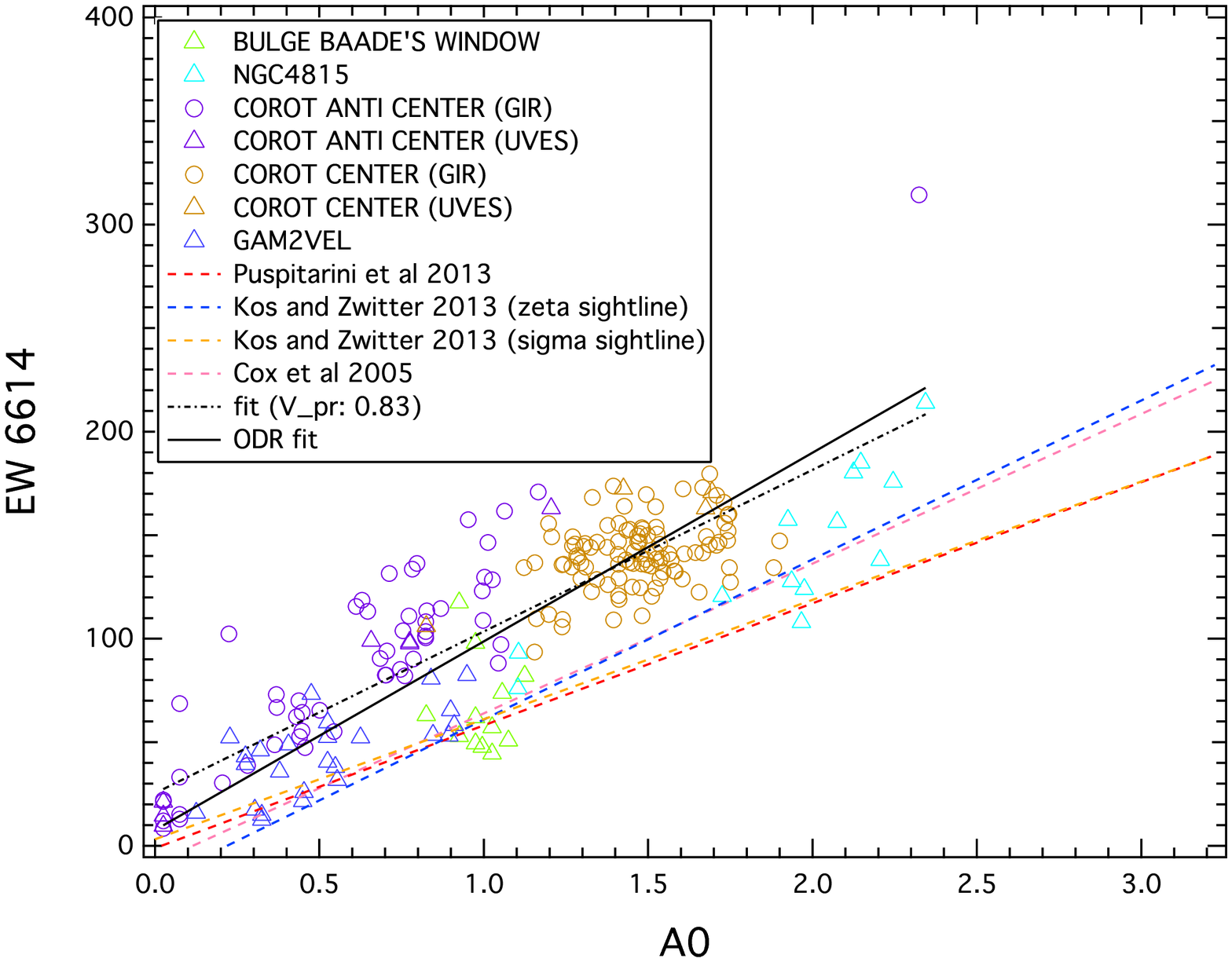}
\caption{DIB vs A0, all fields.}
\label{allfields}
\end{figure}

As discussed in Section 1, a large number of studies were devoted to the correlation between the DIBs and the extinction. Our results provide an opportunity to study further this relation, with for the first time a large selection of DIBs in very different regions of the Galaxy. 
Figure \ref{allfields} shows the whole set of 6283 and 6614 \AA\ DIB EWs as a function of extinction.  
We also display the DIB-extinction relations obtained from previous studies using early-type star data \citep{2013A&A...555A..25P,vos11}.  To convert the color excess values E(B-V) listed in the previous works into extinction values A$_{0}$, we have assumed that  $A_{0}/E(B-V)= 3.2882 + 0.04397  \times E(B-V) $ in all directions. We have fitted the DIB-extinction relationship independently of the error bars, and also using both errors (in extinction and EW) using the orthogonal distance regression method (ODR) \citep{Boggs}. 

We have compared our correlation coefficients with those obtained from previous studies based on early-type target stars, characterized by well known extinctions and excellent spectra. It is remarkable that despite the complexity of the global adjustment and the presence of the stellar lines, the correlation between the 6283 \AA\ DIB and the reddening is found to be tighter, as shown by the error-independent Pearson correlation coefficient of 0.91. Such a value is above most previous determinations, e.g. the Friedman et al (2011) coefficient of 0.82. We believe that our use of late-type stars is the dominant reason for a globally decreased dispersion, 
because we avoid the radiation field effects on the DIB carriers that arise around hot stars. Instead, our LOS cross mainly clouds that are far from those radiation sources. Such a conclusion is in agreement with the results of \citet{chen13}.

In the case of the weaker and narrower 6614 DIB, our correlation coefficient is 0.83, i.e. on the same order than the Friedman et al (2011) coefficient. We believe that here the absence of a correlation increase is due to the strong impact of the residual stellar features. This impact is much stronger than on the 6283 DIB due to the  smaller DIB width, closer to the stellar line width. Better synthetic stellar models should help to reduce this source of uncertainty.

From Fig. \ref{allfields}, we remark that our measured EWs are globally higher than what 
has generally been 
derived from early type stars. This is especially clear for the inner disk and CoRoT anti-center fields and may be explained by the fact that we avoid some of the strong DIB suppressions that arise in the environment of UV-bright stars.  For other sight-lines like the one towards NGC4815 there are no significant differences from the DIB-color excess relations based on early type stars. Finally, we note that more dispersion  seems to be present for the local clouds, which may be explained by averaging effects along large distances.

\subsection{Spatial distribution}
\begin{figure}[htbp]
\begin{center}
\includegraphics[width=\linewidth]{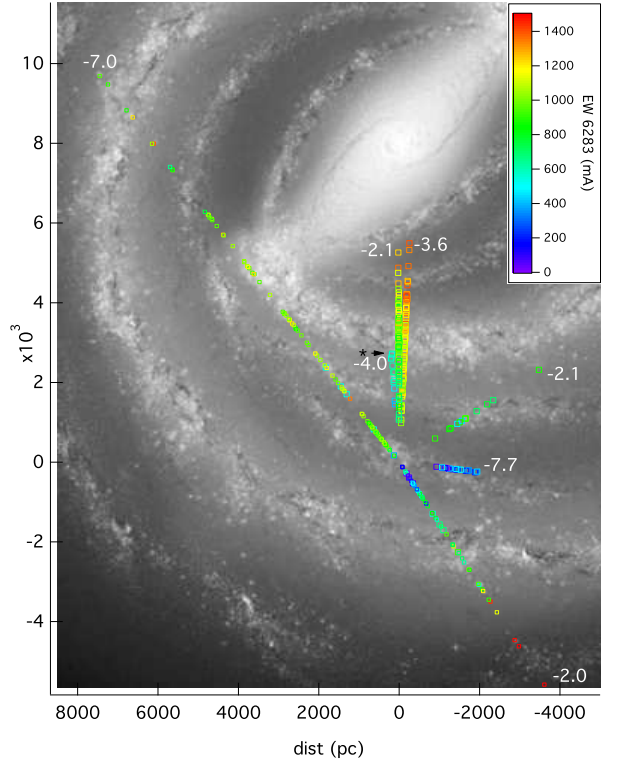}
\caption{Projections of the target stars onto the  face on map of the Galaxy (image from Churchwell et al. 2009). Units are parsecs, counted from the Sun, with d$_{Sun}$ $\simeq$ 8kpc. The color coding corresponds to the equivalent of the 6283 \AA\ DIB, either directly measured, or, when not measured, estimated from the other DIB measurements using the  average EW(6283)/EW(8620) or EW(6283)/EW(6614) ratios computed from the whole dataset. The black asterisk and small arrow mark the Bulge field  5 (Baade Window direction) for which the X coordinate has been multiplied by 4 to avoid confusion with the other directions. The galactic latitude of each field is indicated at the extremity of the sightline.}
\label{gxmap}
\end{center}
\end{figure}

Figure \ref{gxmap} shows the projections of the target stars onto the Galactic plane, superimposed on a face-on map of the Milky Way. The color represents the 6283 DIB EW when it is measured, and a "6283-equivalent" value deduced by simply scaling the 6614 or 8620 DIBs based on the mean 6614/6283 and 8620/6283 ratios. When LOS are about the Plane, the EW reflects the spiral arm crossings. This is no longer the case when the  latitude is increasing, as can be seen in the figure where the latitudes are indicated for each LOS. 

\section{Discussion and perspectives} 


We have developed and applied automated methods of extraction of multi-component DIBs from stellar spectra. These methods can be applied to any kind of stellar spectrum, as long as the stellar parameters are known. In particular it can handle cool star spectra despite their complex continua. 
Here we have presented the results of our automated adjustments when they are applied to ESO/FLAMES high resolution spectra of red giants and F, G, K dwarfs that are part of the GES first data release and about the same number of FLAMES spectra from a previous program about the inner disk. We have extracted three DIBs and studied their strengths and velocity shifts.  
 
The comparison between the DIB strengths and spectro-photometric estimated extinctions reveals a significant correlation and demonstrates that we successfully extract the DIB EWs despite the stellar and telluric absorptions. This correlation suggests that the link between the DIB strength and the extinction does not vary in a large extent among regions of the Galaxy that span galactocentric radii from 2.5 to 13 kpc.  From this dataset we find broad consistency between the DIB distance profiles and the estimated locations/extents of the Local, Perseus 
arms. We also find agreement between the line-of-sight velocity structure deduced from the HI 21 cm emission spectra and the DIBs velocities. This shows that on the large scales DIBs may be a kinematical tool in the same way IS gaseous lines are commonly used. This opens perspectives for the study of the most external arms for which very few measurements of the DIB abundance do exist.

Altogether these results show how DIBs can be used to reconstruct the large-scale distribution of the interstellar medium  in the Galaxy, and may be especially useful for distant clouds because in this case they are strong enough but not saturated. In addition, DIBs in more distant clouds (like Perseus) are found more tightly correlated with the extinction and less spatially/ angularly variable than in the local clouds, which we interpret as an effect of distance-averaging. It also confirms that when using cool (respectively, both cool and distant) stars the effects of a strong radiation field on the DIB abundance and/or ionization are minimized (respectively, both minimized and averaged out), and DIBs follow more closely the extinction. This latter aspect is quantitatively confirmed by the correlation coefficients we obtain when assembling all measurements. In the case of the broad 6283 \AA\ DIB, the Pearson coefficient is 0.91, significantly above previous determinations based on early-type stars despite the extremely large distances between the probed areas and the presence of the stellar lines. This implies that DIBs can be used as a first prior for the extinction in the absence of any other information. However, we note that for one field, the  $\gamma$ Vel cluster direction, there is a more complex relationship between DIBs and extinction estimates. The proximity of the absorber and the presence of bright UV stars is probably responsible for this complexity and the departures from the average conditions. Finally, we note that for two sight-lines the measured DIBs are slightly stronger than previous relationships based on early-type targets have predicted. We believe that this is also due to the absence of strong environmental effects. This deserves further 
study, as the data presented here are still too limited to permit to draw definitive conclusions.

\begin{figure}[h!]
\centering
	\includegraphics[width=0.43\linewidth]{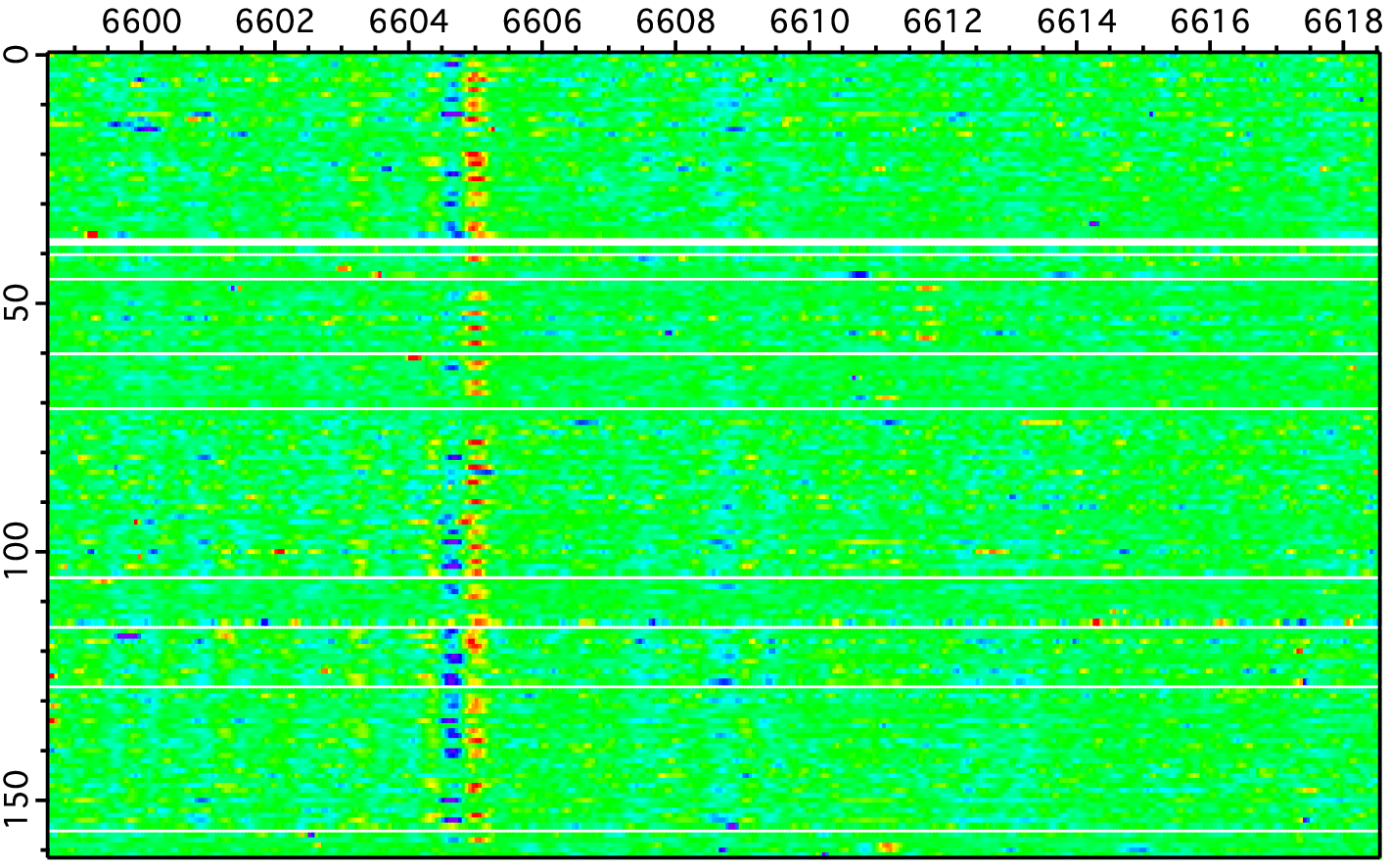}
	\includegraphics[width=0.5\linewidth]{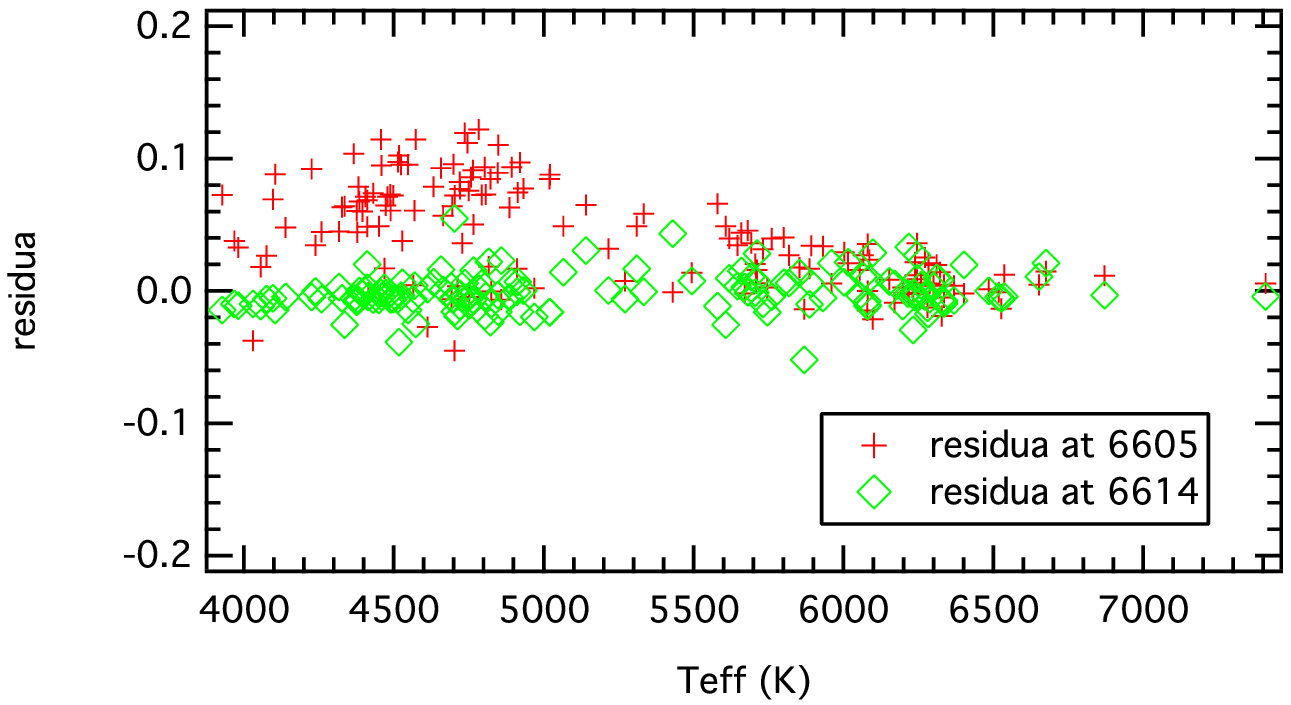}
\caption{Left: Fit residuals (the difference between the data spectrum and the fitted model) for all of target stars in the CoRoT fields observed by GIRAFFE ( 6613.6 \AA\ DIB). The left axis is the star stacking axis and the top axis is the wavelength (after spectra have all been shifted to their respective stellar frames). The color shows the residual intensity.  Zero value (green) corresponds to well-fitted spectra. 
The residual (red and blue for positive and negative residual, respectively) is due to imperfect stellar modeling. Right: The residuals at chosen wavelengths: at $\sim$ 6605 \AA\ (shown in red mark) and at $\sim$ 6614 \AA\ (shown in green). The $\sim$ 6605 \AA\ region shows higher positive residuals due to overestimate of stellar line model in this region. At 6614 \AA, we have fit residuals about 0.
}
\label{residual}%
\end{figure}

\begin{figure}[h!]
\centering
	\includegraphics[width=0.43\linewidth]{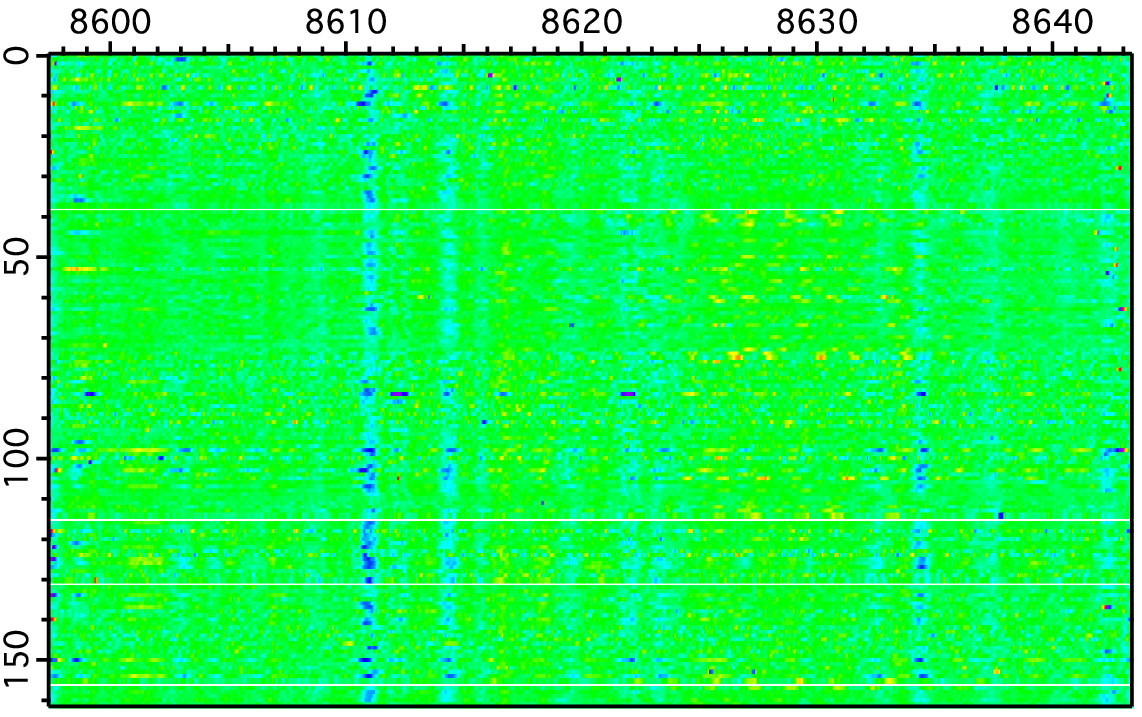}
	\includegraphics[width=0.5\linewidth]{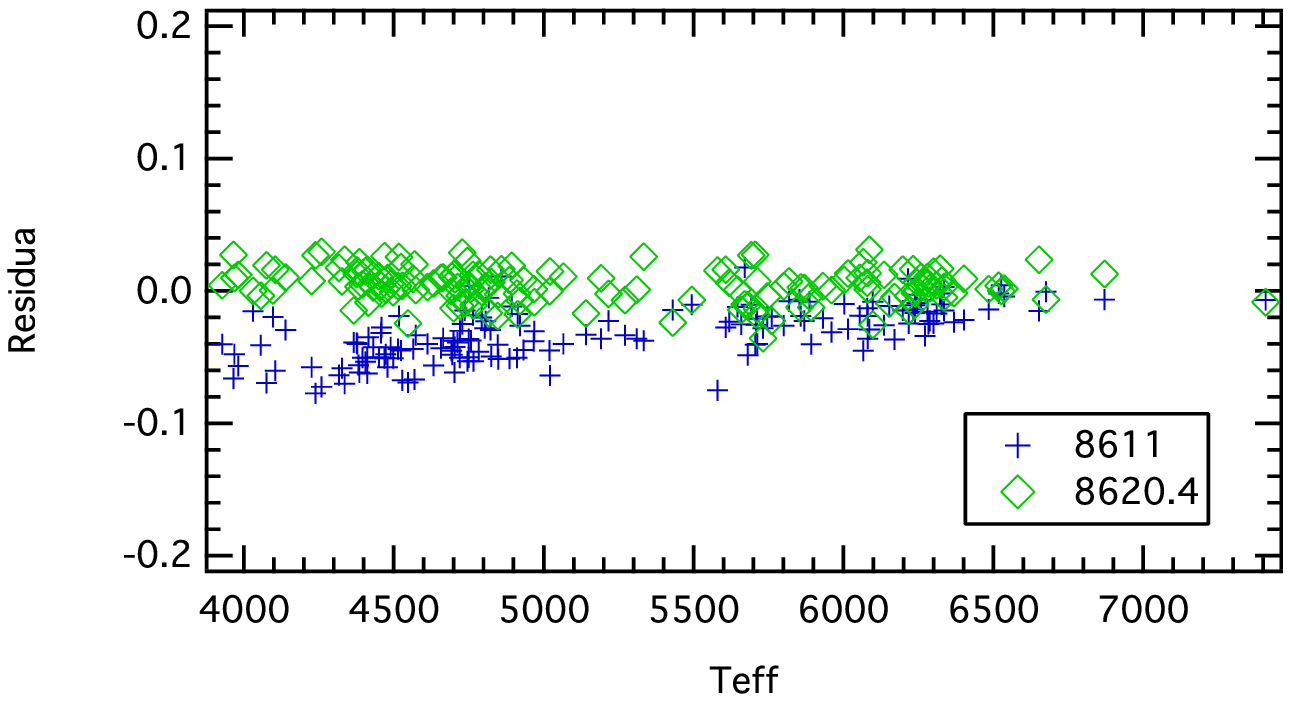}
\caption{Same as Fig. \ref{residual}  but for 8620 DIB. Left: Fit residuals for  all of target stars over wavelength. Right: The residuals at chosen wavelengths. The 8611 \AA\  region shows higher negative residuals due to underestimate od stellar line model.  }
\label{residual2}%
\end{figure}

There is still room for a number of improvements of the synthetic stellar spectrum computations, and subsequent DIB measurements. Here we have used the most probable values of the stellar parameters, and did not allow for any uncertainty. Moreover we did not make any use of individual abundances, although in the case of the GES spectra most of them are determined. One reason is our choice of a homogenous treatment of both GES spectra and other data for which the individual abundance measurements were not available. The second reason is our findings, already developed by \cite{chen13} that the main source of bad quality adjustments of the stellar synthetic spectra to the data is clearly linked to specific spectral lines that are systematically under- or over-estimated, or simply missing (see Fig. \ref{residual}, \ref{residual2}), and allowing for small changes of the parameters would not solve for those discrepancies. Work is in progress to correct for those systematics that must be done before fine tuning of the parameters within the GES error bars or use of individual abundances is done. This should result in a better accuracy of the DIB strength and allow to go further in the kinematical analysis, here still limited to the detection of velocity shifts above 5 to 10 km.s$^{-1}$ depending on DIBs and the signal. Improvements of the fitting strategy are also in progress, in particular the simultaneous adjustment of NaI lines and all measurable DIBs is expected to provide more reliable results. There is also room for an improved strategy regarding the choice of the number of velocity components. We have explored criteria based on the DIB velocity shift, however other methods should also be elaborated and tested. This will be the subject of further studies based on larger datasets. Finally, residuals from sky emission removal are still limiting significantly the DIB extraction in some cases, and special attention must be devoted to this problem.

Globally these results pave the way to three-dimensional mapping of the Galactic ISM based on DIB absorption measurements from current or future stellar spectroscopic surveys. Like all three-dimensional maps, future DIB-based maps are expected to gain in accuracy in a considerable way when Gaia parallax measurements will be available. Finally, as illustrated by the CoRoT anti-center line-of-sight, a very promising aspect that is specific to these DIB spectroscopic measurements is the potential detailed comparison, sightline by sightline, between the distance-limited absorption measurements and emission spectra that trace the gas at all distances. This comparison by means of the radial velocities should bring interesting information on the location of the poorly known dust-poor distant gas in outer parts of the spiral arms.

\begin{acknowledgements}

R.L, L.P., and C.B. acknowledge support from the French National Research Agency (ANR) through the STILISM project. 
L. S. and S. D. acknowledge the support of Sonderforschungsbereich SFB 881  "The Milky Way system " (subprojects A4 and A5) of the German Research Foundation (DFG), and of Project IC120009 "Millennium Institute of Astrophysics (MAS)" of Iniciativa Científica Milenio del Ministerio de Economía, Fomento y Turismo de Chile

This work was partly supported by the European Union FP7 programme through ERC grant number 320360 and by the Leverhulme Trust through grant RPG-2012-541. We acknowledge the support from INAF and Ministero dell' Istruzione, dell' Universit\`a' e della Ricerca (MIUR) in the form of the grant "Premiale VLT 2012". The results presented here benefit from discussions held during the Gaia-ESO workshops and conferences supported by the ESF (European Science Foundation) through the GREAT Research Network Programme.

\end{acknowledgements}


\section{Appendix: Estimate of the error on the DIB equivalent width}

\end{document}